\documentclass{elsarticle}
\usepackage{amsmath,amssymb,graphicx}
\journal{Nuclear Physics A}
\usepackage{graphicx,pstricks}
\usepackage[colorlinks=true]{hyperref}
\usepackage{color}
\usepackage{url}

\newcommand{\bkt}{\boldsymbol{k}_\perp}
\newcommand{\blt}{\boldsymbol{l}_\perp}
\newcommand{\bpt}{\boldsymbol{p}_\perp}
\newcommand{\bqt}{\boldsymbol{q}_\perp}
\newcommand{\bxt}{\boldsymbol{x}_\perp}
\newcommand{\byt}{\boldsymbol{y}_\perp}
\newcommand{\bbt}{\boldsymbol{b}}
\newcommand{\bk}{\boldsymbol{k}}
\newcommand{\bat}{\boldsymbol{a}_\perp}
\newcommand{\br}{\boldsymbol{r}_\perp}
\newcommand{\bu}{\boldsymbol{u}_\perp}
\newcommand{\bv}{\boldsymbol{v}_\perp}
\newcommand{\bw}{\boldsymbol{w}_\perp}
\newcommand{\bx}{\boldsymbol{x}_\perp}

\newcommand{\by}{\boldsymbol{y}_\perp}
\newcommand{\bz}{\boldsymbol{z}_\perp}
\newcommand{\hbu}{\hat{\boldsymbol{u}}_\perp}
\newcommand{\hbp}{\hat{\boldsymbol{p}}_\perp}
\newcommand{\hbr}{\hat{\boldsymbol{r}}_\perp}

\newcommand{\calA}{\mathcal{A}}
\newcommand{\calM}{\mathcal{M}}
\newcommand{\calO}{\mathcal{O}}
\newcommand{\LQCD}{\Lambda_{\rm QCD}}
\newcommand{\Tr}{\text{Tr}}

\newcommand{\be}{\begin{equation}}
\newcommand{\ee}{\end{equation}}
\newcommand{\ba}{\begin{eqnarray}}
\newcommand{\ea}{\end{eqnarray}}

\def\roughly#1{\mathrel{\raise.3ex\hbox{$#1$\kern-.75em%
\lower1ex\hbox{$\sim$}}}}
\def\lsim{\roughly<}
\def\gsim{\roughly>}

\def\slashchar#1{\setbox0=\hbox{$#1$}  
   \dimen0=\wd0     
   \setbox1=\hbox{/} \dimen1=\wd1  
   \ifdim\dimen0>\dimen1   
      \rlap{\hbox to \dimen0{\hfil/\hfil}} 
      #1     
   \else     
      \rlap{\hbox to \dimen1{\hfil$#1$\hfil}} 
      /      
   \fi}

\makeatletter
\def\overbracket#1{\mathop{\vbox{\ialign{##\crcr\noalign{\kern3\p@}
\downbracketfill\crcr\noalign{\kern3\p@\nointerlineskip}
$\hfil\displaystyle{#1}\hfil$\crcr}}}\limits}
\def\underbracket#1{\mathop{\vtop{\ialign{##\crcr
$\hfil\displaystyle{#1}\hfil$\crcr\noalign{\kern3\p@\nointerlineskip}
\upbracketfill\crcr\noalign{\kern3\p@}}}}\limits}
\def\upbracketfill{$\m@th\makesm@sh{\llap{\vrule\@height3\p@\@width.7\p@}}%
\leaders\vrule\@height.7\p@\hfill
\makesm@sh{\rlap{\vrule\@height3\p@\@width.7\p@}}$}
\def\downbracketfill{$\m@th
\makesm@sh{\llap{\vrule\@height.7\p@\@depth2.3\p@\@width.7\p@}}%
\leaders\vrule\@height.7\p@\hfill
\makesm@sh{\rlap{\vrule\@height.7\p@\@depth2.3\p@\@width.7\p@}}$}
\makeatother

\begin{document}
\begin{frontmatter}

\title{Photon from the annihilation process with CGC in the pA collision}
\author[zagreb,tokyo]{Sanjin Beni\' c}
\author[tokyo]{Kenji Fukushima}
\address[zagreb]{Physics Department, Faculty of Science,
                 University of Zagreb, Zagreb 10000, Croatia}
\address[tokyo]{Department of Physics, The University of Tokyo,
                7-3-1 Hongo, Bunkyo-ku, Tokyo 113-0033, Japan}
\begin{abstract}
 We discuss the photon production in the $pA$ collision in a framework
 of the color glass condensate (CGC) with expansion in terms of the
 proton color source $\rho_p$.  We work in a regime where the color
 density $\rho_A$ of the nucleus is large enough to justify the CGC
 treatment, while soft gluons in the proton could be dominant over
 quark components but do not yet belong to the CGC regime, so that we
 can still expand the amplitude in powers of $\rho_p$.  The
 zeroth-order contribution to the photon production is known to appear
 from the Bremsstrahlung process and the first-order corrections
 consist of the Bremsstrahlung diagrams with pair produced quarks and
 the annihilation diagrams of quarks involving a gluon sourced by
 $\rho_p$.  Because the final states are different there is no
 interference between these two processes.  In this work we elucidate
 calculation procedures in details focusing on the annihilation
 diagrams only.  Using the McLerran-Venugopalan model for the color
 average we numerically calculate the photon production rate and
 discuss functional forms that fit the numerical results.
\end{abstract}


\end{frontmatter}

\section{Introduction}

Color glass condensate (CGC) is a well-developed theoretical framework
in which perturbative expansion works at weak coupling
$\alpha_s=g^2/(4\pi)\ll 1$ but large gluon amplitude (occupation
number) $A^\mu\sim \calO(g^{-1})$ requires resummation to take full
account of non-linearity with respect to $g A^\mu\sim \calO(g^0)$.
Such a treatment amounts to the perturbative expansion around a CGC
background field given by a solution of the classical Yang-Mills
equations~\cite{McLerran:1993ni}. In the CGC regime soft physical
quantities are all characterized by a unique scale called the
saturation momentum $Q_s$.  From the geometrical scaling in the deep
inelastic scattering (DIS), $Q_s$ as a function of Bjorken's $x$ can
be determined experimentally~\cite{Stasto:2000er}.
It is believed that CGC should give a good theoretical description of
the initial dynamics in relativistic heavy-ion
collisions~\cite{Fukushima:2011ca,Gelis:2014tda}.
The CGC computation is also successful in quantitative estimate
of particle production especially for the forward (or backward)
rapidity region and/or for the $pA$ or $dA$
collisions~\cite{Albacete:2014fwa}.  In such cases one could access
smaller $x$ than mid-rapidity region in the $AA$ collision, namely,
$\lesssim 10^{-2}$ for Relativistic Heavy Ion Collider
(RHIC) and $\lesssim 10^{-3}$ for Large Hadron Collider (LHC), which
should make the CGC work better.

Photon is a transparent probe conveying the information on the early
stage of the heavy-ion collision.  It has been observed that the
direct photon spectrum in the $AA$ collision shows thermal exponential
behavior, from which the initial temperature or the slope parameter
has been extracted~\cite{Adare:2008ab} (see Ref.~\cite{Klasen:2013mga} for discussions on robustness with uncertainties in parton distribution functions).  The physical interpretation
of such a slope parameter is, however, sometimes under discussion. For example, Ref.~\cite{Klein-Bosing:2014uaa} found geometrical scaling in the experimental photon spectrum.
There might indeed be some initial-state mechanism that allows for a
thermal-like spectrum, which was assumed as an Ansatz for glasma
photons~\cite{McLerran:2014hza}. To help our theoretical
understanding, the direct CGC calculation of the photon production
rate should be useful.  We should emphasize that the photon estimates
from a thermalized quark-gluon plasma~\cite{Baier:1991em} and from
hadronic matter~\cite{Heffernan:2014mla} have been somehow
established, and in this sense, the CGC photon is the last missing
piece and is an urgent problem to be solved.  The CGC photon has been
considered in some pioneering works in the $pA$
case~\cite{Gelis:2002ki} at the lowest order and the
$AA$ case~\cite{Tanji:2015ata} as well.  Here, the lowest order means
the zeroth order in the expansion in powers of the proton color
density $\rho_p$, as formulated first for the gluon production
problem~\cite{Dumitru:2001ux,Blaizot:2004wu,Gelis:2005pt,Fukushima:2008ya}
and later extended to the quark production~\cite{Blaizot:2004wv}.

At sufficiently high collision energy, even in the $pA$ collision,
some shape of ``matter'' like the quark-gluon plasma may be created,
or more precisely speaking, there should be an onset of
\textit{collectivity} even for small systems such as $pA$, $dA$, and
even $pp$.  While RHIC data for the photon spectrum in $d+$Au at
$\sqrt{s_{NN}} = 200\;\text{GeV}$~\cite{Adare:2012vn} are fairly
consistent with rescaled perturbative QCD (pQCD) results, the
situation is not conclusive yet.  For example, the RHIC data could
accommodate thermal photons on top of pQCD as discussed in
Ref.~\cite{Shen:2015qba}.  Anticipating 
forthcoming LHC/RHIC data for the photon spectrum in $p+$Pb and
$p+$Au, at $\sqrt{s_{NN}}=5.02\;\text{TeV}$~\cite{LHC}
a quantitative prediction from the CGC fields is definitely needed.
For this purpose, because $\rho_p$ goes up with energy, the
first-order corrections of $\calO(\rho_p^2)$ (in the rate, and
$\calO(\rho_p)$ in the amplitude) should be of increasing
importance.  As we will argue later, actually, we can
even consider a \textit{semi-CGC regime} for a systematic treatment,
in which the first-order terms can be comparable to the zeroth-order
ones, while $\rho_p$ is still dilute enough to validate a systematic
expansion in powers of $\rho_p$.  This observation clearly motivates
us to take a careful look at the first-order contributions to the
photon production in the $pA$ case.

This paper is organized as follows. In Sec.~\ref{sec:processes} we
make a classification of the zeroth- and the first-order diagrams
contributing to the photon production in the $pA$ collision.
In Sec.~\ref{sec:ampl} we analytically calculate the amplitude for the
annihilation diagram (to be precisely defined below) with the main
result given by Eq.~\eqref{eq:ampl2}.
The following Sec.~\ref{sec:rat} is devoted to a calculation of the
photon production rate and the main result is found in
Eq.~\eqref{eq:ratf}.
The numerical computation of the rate is reported in
Sec.~\ref{sec:num}.
Conclusions are made in the final Sec.~\ref{sec:conc}.
Technical details about derivations of some key equations used in the
paper are collected in the Appendices.

\section{Zeroth and first-order diagrams}
\label{sec:processes}

In the CGC framework the collision of the proton $p$ (light
projectile) and the nucleus $A$ (dense target) at high energy is
dominated by classical color fields representing the small-$x$
partons.  For definiteness, we will take the nucleus to be moving
along $x^-$ and the proton along $x^+$, where $x^\pm=(x^0\pm
x^3)/\sqrt{2}$.  We postulate that the nucleus color density $\rho_A$
is dense as $\rho_A\sim\calO(g^{-2})$ (to make the gauge field of
$\calO(g^{-1})$ in our convention) and the color density of the
projectile $\rho_p$ is less dense as
$\calO(g^{-1})<\rho_p< \calO(g^{-2})$.

\begin{figure}
 \begin{minipage}{0.25\textwidth}
  \begin{center}
   {\large (a)}
   \includegraphics[width=\textwidth]{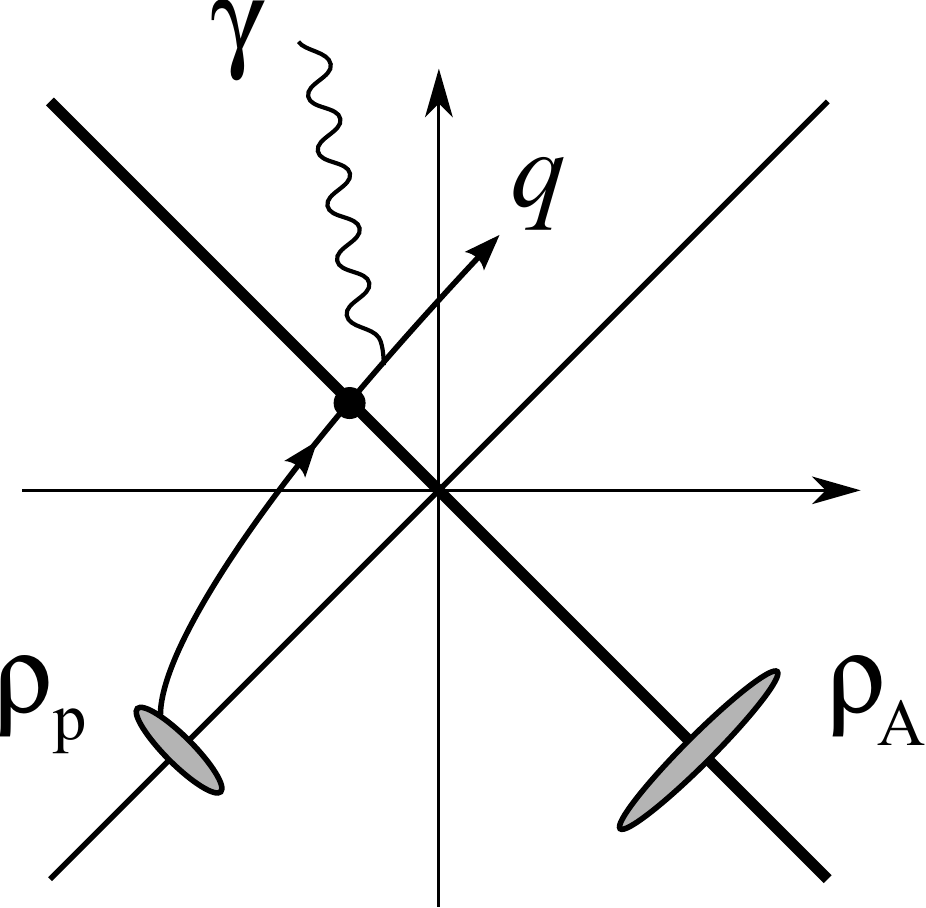}
  \end{center}
 \end{minipage}
 \hspace{0.1\textwidth}
 \begin{minipage}{0.25\textwidth}
  \begin{center}
   {\large (b)}
   \includegraphics[width=\textwidth]{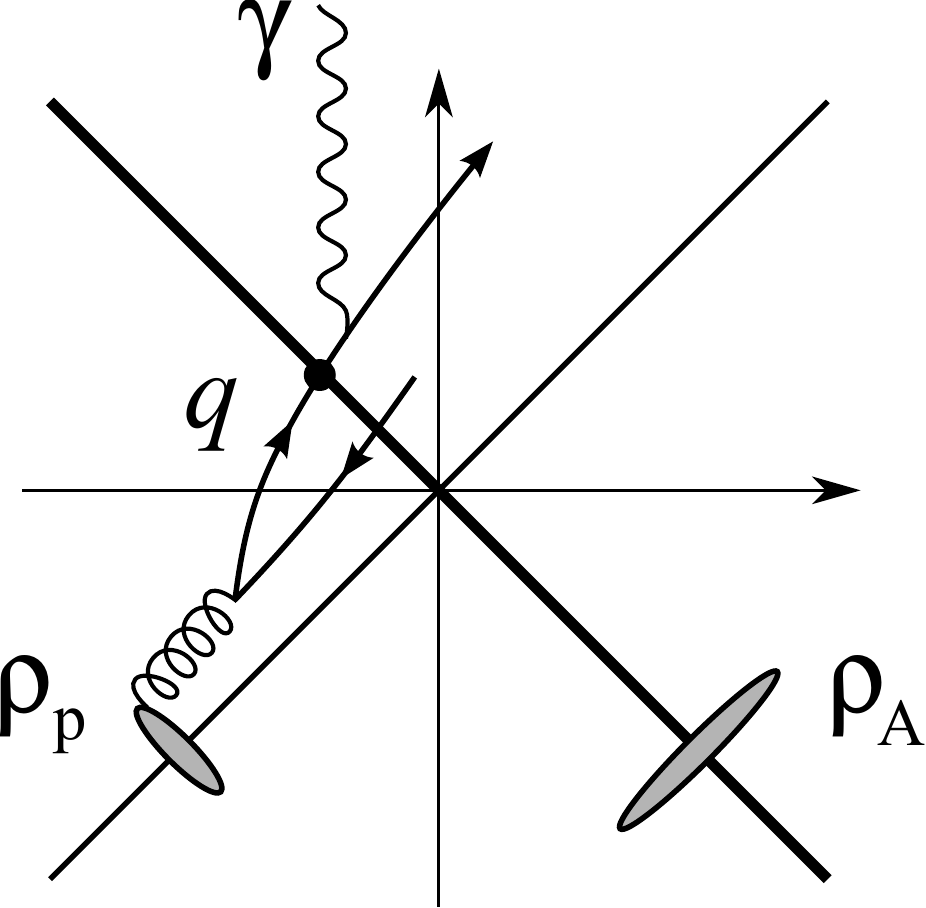}
  \end{center}
 \end{minipage}
 \hspace{0.1\textwidth}
 \begin{minipage}{0.25\textwidth}
  \begin{center}
   {\large (c)}
   \includegraphics[width=\textwidth]{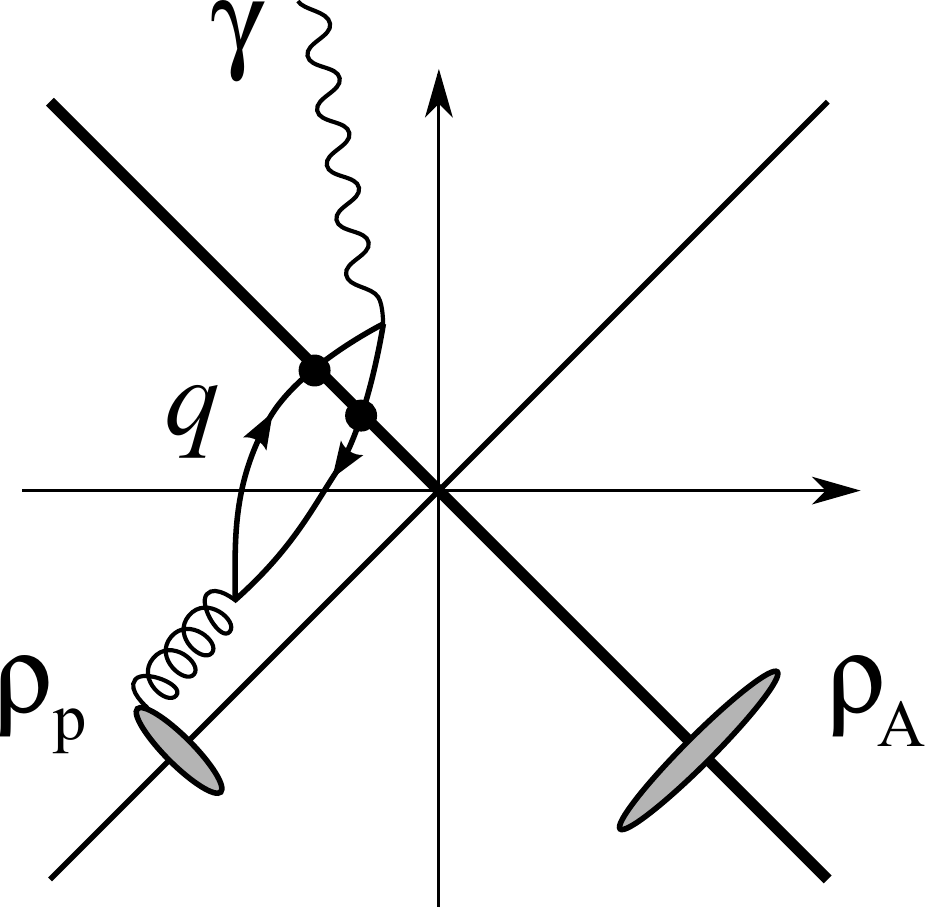}
  \end{center}
 \end{minipage}
 \caption{Diagrams contributing to the photon production in the $pA$
   collision.  The solid line, the curly line, and the wavy line
   represent quarks, gluons, and photons, respectively.  Blobs on
   the $x^-$ axis represent CGC resummed gluonic interaction with
   $\rho_A$ (see also Fig.~\ref{fig:vertex}).  (a) Zeroth-order
   Bremsstrahlung process (considered in Ref.~\cite{Gelis:2002ki}).
   (b) First-order Bremsstrahlung process that can take over (a) for
   abundant gluons in $p$ in the semi-CGC regime.
   (c) First-order annihilation process that can be also of the same
   order as (b) in the semi-CGC regime (considered in this work).
   Some other diagrams such as (a) with the photon line attached to
   the quark in the region $x^+<0$ (see Fig.~\ref{fig:vertex}), (b)
   with the photon coupled to the antiquark are not shown just for
   simplicity.  (c) may have four different combinations depending on
   the relative position of the photon and the gluon vertices, among
   which only one shown in (c) is non-zero (see Sec.~\ref{sec:ampl}
   for more discussions).}
 \label{fig:LO}
\end{figure}

\begin{figure}
  \begin{center}
  \includegraphics[width=0.7\textwidth]{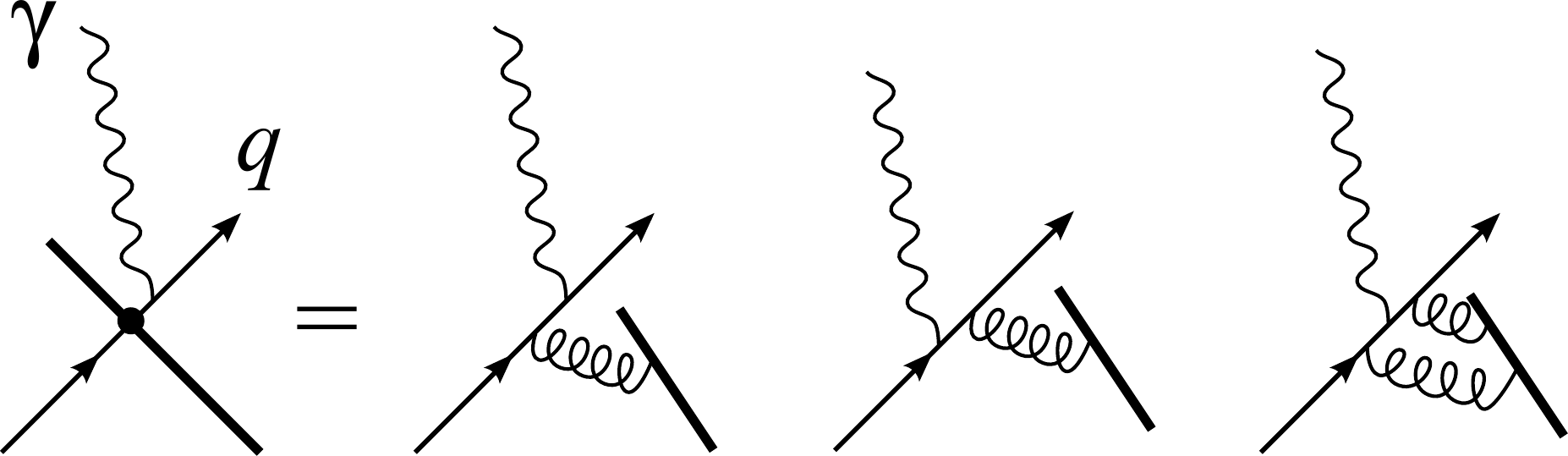}
  \end{center}
  \caption{Three distinct processes of photon emitted before, after,
    and during gluon scatterings.}
  \label{fig:vertex}
\end{figure}

For the photon production with CGC, the zeroth-order contribution in
the $pA$ collision $\sim\calO(\rho_p^0)$ is the Bremsstrahlung process
as shown in Fig.~\ref{fig:LO}~(a)~\cite{Gelis:2002ki}, which is
actually a CGC generalization of the Compton scattering that would be
the leading-order contribution in the hard thermal loop calculation.
For the Bremsstrahlung process, the quark that emits a photon should
interact with gluons, and such gluon scatterings with CGC
(represented by a blob on the $x^-$ axis) are \textit{not} suppressed
by the strong coupling constant compensated by the CGC fields.  This
zeroth-order diagram gives a photon rate of $\calO(\alpha n_q)$ with
the fine structure constant $\alpha=e^2/(4\pi)$ and the quark number
density $n_q$ in $p$.  Here, we note that the blob in
Fig.~\ref{fig:LO} is a bit sloppy representation of physical processes
and it actually contains three distinct contributions; the first one
with photon emitted before gluon scatterings, the second one with
photon emitted after gluon scatterings, and the last one with photon
emitted during gluons scatterings, as illustrated in
Fig.~\ref{fig:vertex}.  As argued in Ref.~\cite{Gelis:2002ki}, the
last diagram is vanishing in the limit of a fast-moving projectile at
the speed of the light.

In the semi-CGC regime, we may ideally think that we can neglect
valence and sea quarks in $p$, and then the dominating gluons can
couple to photons only through quarks.  Therefore, in the first order,
the Bremsstrahlung process is possible only from pair produced quarks
as in Fig.~\ref{fig:LO}~(b), which is of $\calO(\alpha \alpha_s n_g)$
with $n_g$ representing the gluon number density in $p$.  In reality
we should say that the first-order contribution as shown in
Fig.~\ref{fig:LO}~(a) is not practically small but just comparable
to Fig.~\ref{fig:LO}~(b) even though gluons are such abundant;  we
know that around $x\sim 10^{-2}$ at RHIC the gluon density $n_g$ is
one order of magnitude larger than $n_q$ and so $\alpha_s n_g$ with
$\alpha_s\sim \calO(10^{-1})$ cannot completely supersede $n_q$, while
the first-order terms would be more dominant at LHC.

There is another important diagram of the same first order, which
is shown in Fig.~\ref{fig:LO}~(c).  This represents the annihilation
process giving a rate of $\calO(\alpha \alpha_s n_g)$.  One might think
that there should be also a zeroth-order annihilation process
involving two quarks in $p$ in the way similar to
Fig.~\ref{fig:LO}~(a), but such a process would result in
$\calO(\alpha n_q^2)$ and so it is suppressed by one more $n_q$.  In
conclusion, the diagrams (a) at the zeroth order, (b) and (c) at the
first order in the $\rho_p$ expansion are physically of the same order
in the semi-CGC regime.  The diagram (a) was computed in
Ref.~\cite{Gelis:2002ki} and so the remaining task is to compute
diagrams (b) and (c).  The diagram (b) can be partially considered as
a correction to the diagram (a) once the integration over the
anti-quark phase-space is performed.  Nevertheless, (b) also yields a
contribution kinematically separate from diagram (a), which we will
discuss in details in separate publication.

In this work we will focus only on the process of
Fig.~\ref{fig:LO}~(c).  Although this is a part of the whole
contributions, it is conceivable that (a) and (b) may become more
relevant for soft photons with momenta $\lesssim Q_s$ and (c) would
be more dominating for hard photons with momenta $\sim Q_s$.  This is
because soft photons are enhanced in Figs.~\ref{fig:LO}~(a) and (b)
with collinear enhancement and the momentum $\sim Q_s$ provided by the
interaction with $\rho_A$ can be taken away mostly by quarks.  In
contrast, in Fig.~\ref{fig:LO}~(c), quark momenta all go to the
produced photon, and naturally, the emitted photon should carry
momenta $\sim Q_s$.  We here point out that diagrams (b) and (c) have
a different final state, and so their rates (not amplitudes) can be
computed individually.  As a final remark, we note that that the
quark-loop contributions from Fig.~\ref{fig:LO}~(c) are suppressed by
the charge cancellation among three flavors, but the three-flavor
symmetry is broken by $s$-quark in the soft sector and by $c$-quark in
the hard sector.  For phenomenological applications, one should keep
this in mind, though phenomenological discussions are beyond our
current scope in this present work.

\section{Calculation of the amplitude}
\label{sec:ampl}

We now proceed to the concrete calculation of the process in
Fig.~\ref{fig:LO}~(c).  As a prerequisite to our work and for readers'
convenience we first summarize the already established results in the
CGC framework on the $\calO(\rho_p^1)$ gluon field and the
$\calO(\rho_p^0)$ quark propagator.

\subsection{Classical gluon fields}

The starting point in the CGC-based calculation is the solution of the
Yang-Mills equations, $[D_\mu,\mathcal{F}^{\mu\nu}] = J^\nu$, in the
presence of the current:
\be
 J^\nu(x) = g\delta^{\nu+}\delta(x^-) \rho_p(\bxt)
  + g\delta^{\nu-}\delta(x^+) \rho_A(\bxt)
\ee
with sources $\rho_p$ and $\rho_A$ localized on the light cone.
Provided that $\rho_p \ll \rho_A$ the classical gluon field
$\mathcal{A}^\mu(x)$ is solved from the Yang-Mills equations order by
order in $\rho_p$.  We denote
$\calA^\mu = \calA_{(0)}^\mu + \calA_{(1)}^\mu$ where
$\calA_{(0)}^\mu$ and $\calA_{(1)}^\mu$ are of zeroth and first order
in terms of $\rho_p$.  Throughout this work we will perform
calculations in the light-cone gauge, i.e.\ $\calA^+ = 0$, with which
the gluon field of a single nuclei (with full resummation in $\rho_A$)
was first derived in Ref.~\cite{Kovchegov:1996ty}.  The $O(\rho_p^1)$
correction was calculated for the first time in
Ref.~\cite{Blaizot:2004wu}.  For the calculation in other gauges, see
Refs.~\cite{Dumitru:2001ux,Gelis:2005pt,Fukushima:2008ya}.

In the covariant gauge the gluon field $\mathcal{A}_{(0)}^\mu(x)$ is
given simply by a solution of the Poisson equation as
\be
 \calA_{(0)}^\mu(x) = -g n^\mu\delta(x^+)\frac{1}{\partial_\perp^2}
 \rho_A(\bxt)\;,
\label{eq:A0}
\ee
where $n^\mu = \delta^{\mu -}$.  This is a covariant gauge solution
for the nucleus, but above $\calA_{(0)}^\mu$ is consistent with the
light-cone gauge for the proton, which is a theoretical trick to
simplify the $pA$ calculation
significantly~\cite{Gelis:2005pt,Fukushima:2008ya}.  We note that
$\calA_{(0)}^\mu$ does not depend on $x^-$ because of time dilatation,
and proportional to $\delta(x^+)$ in the limit of Lorentz contraction.
The higher-order gluon field $\mathcal{A}_{(1)}^\mu(x)$ has a
different functional form in two regions, $x^+<0$ and $x^+>0$, and we
will use the notation of Ref.~\cite{Fukushima:2008ya} to denote them
as $\calA_{(1)}^\mu(x) = \calA_{(1<)}^\mu(x) + \calA_{(1>)}^\mu(x)$.
In momentum space with Fourier transformation,
\be
 \calA_{(1)}^\mu(p) = \int d^4 x\, e^{ip\cdot x} \calA_{(1)}^\mu(x) \;,
\ee
we can find explicit forms of $\calA_{(1)}^\mu(x)$ as follows.  In the
region $x^+<0$, the gluon has no interaction with the CGC yet, and
thus we have \cite{Gelis:2005pt,Fukushima:2008ya}
\be
\begin{split}
 & \calA_{(1<)}^+(p) = 0 \;,\qquad \calA_{(1<)}^-(p) = 0 \;,\\
 & -p^2 \calA_{(1<)}^i(p) = -ig p^i
 \frac{p^2}{(p^+ + i\epsilon)(p^- - i\epsilon)}
 \frac{\rho_p(\bpt)}{p_\perp^2} \;,
\label{eq:gl1}
\end{split}
\ee
which have no dependence on $\rho_A$.  Proceeding to the region
$x^+>0$ the gluon field picks up the adjoint Wilson line associated
with the CGC as
\be
 V(\bx) \equiv \mathcal{P}_{x^+}
 \exp\biggl[ ig\int_{-\infty}^\infty dx^+ \, \calA^{- a}_{(0)}(x) \,
   T_A^a\biggr]\;,
\ee
where $T_A^a$ belong to the adjoint $su(3)$ algebra.  Using this
matrix in transverse momentum space transformed by
\be
 V(\bqt) = \int d^2 x_\perp \, e^{-i\bqt\cdot \bxt} V(\bxt)\;,
\ee
we can give the explicit expression for the field as
\cite{Gelis:2005pt,Fukushima:2008ya} 
\be
 -p^2 \calA_{(1>)}^\mu(p) = -ig\int\frac{d^2 \bqt}{(2\pi)^2}\,
 C^\mu(p;\bqt,\bpt-\bqt)V(\bpt-\bqt)\frac{\rho_p(\bqt)}{q_\perp^2}\;,
\label{eq:A1l}
\ee
where we defined $C^\mu(p;\bqt,\bpt-\bqt)$ as
\be
\begin{split}
 & C^+(p;\bqt,\bpt-\bqt) = 0\;,\\
 & C^-(p;\bqt,\bpt-\bqt)
  = \frac{-2 \bqt \cdot(\bpt - \bqt)}{p^+ + i\epsilon}\;,\\
 & C^i(p;\bqt,\bpt-\bqt)
  = \frac{p^i \,q_\perp^2}{(p^+ +i\epsilon)(p^- +i\epsilon)} - 2 q^i\;.
\end{split}
\ee

\subsection{Quark propagator}

The fact that $\calA_{(0)}^\mu(x)$ is localized at $x^+=0$ makes it
easier to write an analytical expression down for the quark propagator
in the presence of $\calA_{(0)}^\mu(x)$ background.  This result was
established some time ago in Ref.~\cite{Baltz:1998zb} to be
\begin{equation}
 \begin{split}
  & S_{(0)}(x,y) \equiv
   -i\langle\Omega_{\rm out}|T \psi(x)\bar{\psi}(y)|\Omega_{\rm in}\rangle
   = S_F(x-y) \\
  &\qquad\quad +i\theta(x^+)\theta(-y^+)\int d^4 z\,
   \delta(z^+)\bigl[U(\bz)-1\bigr] S_F(x-z)\slashchar{n}S_F(z-y)\\ 
  &\qquad\quad -i\theta(-x^+)\theta(y^+)\int d^4 z\,
   \delta(z^+)\bigl[U^\dag(\bz)-1\bigr]S_F(x-z)\slashchar{n}S_F(z-y)~,
 \end{split}
\label{eq:qprop}
\end{equation}
where the fundamental Wilson line takes care of the multiple
interaction with the CGC gluons, i.e.
\begin{equation}
 U(\bx) \equiv \mathcal{P}_{x^+}
  \exp\biggl[ ig\int_{-\infty}^\infty dx^+ \, \calA^{- a}_{(0)}(x)
  \, T_F^a\biggr]\;,
\end{equation}
and $T_F^a$ belong to the fundamental $su(3)$ algebra.  Here,
$S_F(x-y)$ represents the free Feynman propagator given in a standard
form as
\be
 S_F(x-y) = \int \frac{d^4p}{(2\pi)^4}S_F(p)e^{-ip\cdot(x-y)}\;,
 \qquad
 S_F(p) = \frac{\slashchar{p}+m}{p^2-m^2+i\epsilon}\;.
\label{eq:pfey}
\ee
For the purpose of calculating the photon production rate, it is
useful to decompose the propagator $S_{(0)}(x,y)$ into a direct sum of
four contributions depending on the signs of $x^+$ and $y^+$, i.e.
\be
 S_{(0)}(x,y) = S_{(0>>)}(x,y)+S_{(0><)}(x,y)+S_{(0<>)}(x,y)+S_{(0<<)}(x,y)\;,
\label{eq:s0spl}
\ee
where we defined,
\begin{align}
 S_{(0>>)}(x,y) &\equiv \theta(x^+)\theta(y^+)S_F(x-y)\;,\\
 S_{(0><)}(x,y) &\equiv i\theta(x^+)\theta(-y^+)
  \int d^4 z\, \delta(z^+)U(\bz)S_F(x-z)\slashchar{n}S_F(z-y)\;,\\
 S_{(0<>)}(x,y) &\equiv -i\theta(-x^+)\theta(y^+)\!
  \int d^4 z\, \delta(z^+)U^\dag(\bz)S_F(x\!-\!z)
  \slashchar{n}S_F(z\!-\!y)\;,\\
 S_{(0<<)}(x,y) &\equiv \theta(-x^+)\theta(-y^+)S_F(x-y)\;.
\label{eq:spliprop}
\end{align}
The intuitive meaning of the above decomposition is clear.  For
$S_{(0>>)}$ and $S_{(0<<)}$ there is no crossing with the nucleus CGC
field, while $S_{(0><)}$ and $S_{(0<>)}$ have one crossing at $z$
which should be integrated.
We note that, in what follows, we will sometimes use the well-known
properties of the propagator in the light-cone coordinates.  As is
clear from an explicit manipulation,
\begin{equation}
  \theta(x^+)S_F(x)
  = -i\theta(x^+) \int_{p^+>0}\frac{dp^+ d^2\bpt}{(2\pi)^3}\,
  e^{-ip^+ x^--i\tilde{p}^- x^+ +i\bpt\cdot\bxt}
  \frac{\tilde{\slashchar{p}}+m}{2p^+}\;,
\label{eq:propagation}
\end{equation}
where $\tilde{p}^-=(p_\perp^2+m^2)/(2p^+)-i\epsilon/(2p^+)$, we see
that the particle with $p^\pm > 0$ should propagate in the direction
of increasing $x^+$, while the anti-particle with $p^\pm < 0$ should
propagate in the direction of decreasing $x^+$.

\subsection{Amplitude}
\label{ssec:ampl}

We now give the amplitude from the vacuum $|\Omega_{\rm in}\rangle$ to
a single photon state $|\bk,\lambda\rangle$ with momentum $\bk$ and
polarization $\lambda$ using the LSZ reduction formula.  Expanding in
powers of $\rho_p$ we find,
\be
\begin{split}
 & \calM_\lambda(\bk) \equiv \langle \bk,\lambda|\Omega_{\rm in}\rangle
  = -ie \int d^4 x\, e^{ik\cdot x}\, \langle \Omega_{\rm out}|
  \bar{\psi}(x)\slashchar{\epsilon}_\lambda(\bk) \psi(x)
  |\Omega_{\rm in} \rangle\\
 &\qquad + e g \int d^4 x\, d^4 y\, e^{ik\cdot x}\,
  \langle \Omega_{\rm out} |T \bar{\psi}(x)
  \slashchar{\epsilon}_\lambda(\bk) \psi(x)\bar{\psi}(y)
  \slashchar{\calA}_{(1)}(y) \psi(y)|\Omega_{\rm in} \rangle\;,
\end{split}
\ee
where $\epsilon^\mu_\lambda(\bk)$ is the photon polarization vector.
We will consider the photon in the light-cone gauge so that
$n\cdot \epsilon_\lambda(\bk) = 0$ in addition to the transversality
condition $k \cdot \epsilon_\lambda(\bk) = 0$. 

The above matrix elements are to be evaluated in a $\calA_{(0)}^\mu$
background.  Using the Feynman rules we write them in terms of the
quark propagator~\eqref{eq:qprop} leading to
\begin{equation}
\begin{split}
 \calM_\lambda(\bk) &= -e \int d^4 x\, e^{ik\cdot x}\,
  \Tr\bigl[\slashchar{\epsilon}_\lambda(\bk) S_{(0)}(x,x)\bigr] \\
 &\quad -eg \int d^4 x\, d^4 y\,  e^{ik\cdot x} \, \Tr\bigl[
  \slashchar{\epsilon}_\lambda(\bk) S_{(0)}(x,x)\bigr]\,
  \Tr\bigl[\slashchar{\calA}_{(1)}(y) S_{(0)}(y,y)\bigr]\\
 &\quad +eg \int d^4 x\, d^4 y\, e^{ik\cdot x} \, \Tr\bigl[
  \slashchar{\epsilon}_\lambda(\bk) S_{(0)}(x,y)
  \slashchar{\calA}_{(1)}(y) S_{(0)}(y,x)\bigr]\;,
\label{eq:ampl}
\end{split}
\end{equation}
where we take the trace with respect to the Dirac and the color
indices.  In the above formula, the first term is of $\calO(\rho_p^0)$
while the second (disconnected) and the third (connected)
contributions are of $\calO(\rho_p^1)$.  The formula explicitly
includes the multiple scattering effects through the classical gluon
field $\mathcal{A}_{(1)}$ and the quark propagator $S_{(0)}$.

The $\calO(\rho_p^0)$ amplitude naturally vanishes because no photon
emission occurs without the nucleus interaction.  We can explicitly
demonstrate this by inserting~Eq.~\eqref{eq:qprop} into the first term
of Eq.~\eqref{eq:ampl} as
\begin{align}
 & \int d^4 x\, e^{ik\cdot x}\, \Tr\bigl[
  \slashchar{\epsilon}_\lambda(\bk) S_{(0)}(x,x)\bigr] = (2\pi)^4
  \delta^{(4)}(k) N_c \int \frac{d^4 q}{(2\pi)^4} \Tr_D
  \bigl[\slashchar{\epsilon}_\lambda(\bk) S_F(q)\bigr] \notag\\
 &\qquad\qquad +2\pi i \, \delta(k^+) \Tr_c\bigl[U(\bkt) - (2\pi)^2
    \delta^{(2)}(\bkt)\bigr] \notag\\
  &\qquad\qquad\qquad\times \int \frac{d^4 q}{(2\pi)^4}
  \theta(q^+) \Tr_D\bigl[\slashchar{\epsilon}_\lambda(\bk)
  S_F(q)\slashchar{n}S_F(-k+q)\bigr] \notag\\
 &\qquad\qquad -2\pi i \, \delta(k^+) \Tr_c\bigl[U^\dag(-\bkt) - (2\pi)^2
    \delta^{(2)}(\bkt)\bigr] \notag\\
  &\qquad\qquad\qquad\times \int \frac{d^4 q}{(2\pi)^4}
  \theta(-q^+) \Tr_D\bigl[\slashchar{\epsilon}_\lambda(\bk)
  S_F(q)\slashchar{n}S_F(-k+q)\bigr]\;,
\label{eq:azer}
\end{align}
where we explicitly distinguished the Dirac ($\Tr_D$) and the color
($\Tr_c$) traces.  Due to the $\delta$-function constraints none of
these three terms in Eq.~\eqref{eq:azer} can be kinematically allowed
for an on-shell photon at $\bk\neq 0$, and so the $O(\rho_p^0)$
amplitude vanishes\footnote{Another way to argue that the amplitude at
  this order is zero is from Lorentz covariance.  The only
  four-vectors at disposal are $k^\mu$ and $n^\mu$ (i.e.\ the
  direction of $\calA_{(0)}^\mu$).  By covariance we can write down
  the fermion loop as a linear combination of these two vectors.
  Therefore, when we contract it with the polarization vector, the
  result is zero because $n\cdot \epsilon_\lambda = 0$ and
  $k \cdot \epsilon_\lambda = 0$.}.
We can also show that the second (disconnected) $O(\rho_p^1)$
amplitude in Eq.~\eqref{eq:ampl} vanishes for a similar reason.

\begin{figure}[t]
 \begin{center}
 \begin{minipage}{0.3\textwidth}
  \begin{center}
   {\large (a)}
   \includegraphics[width=\textwidth]{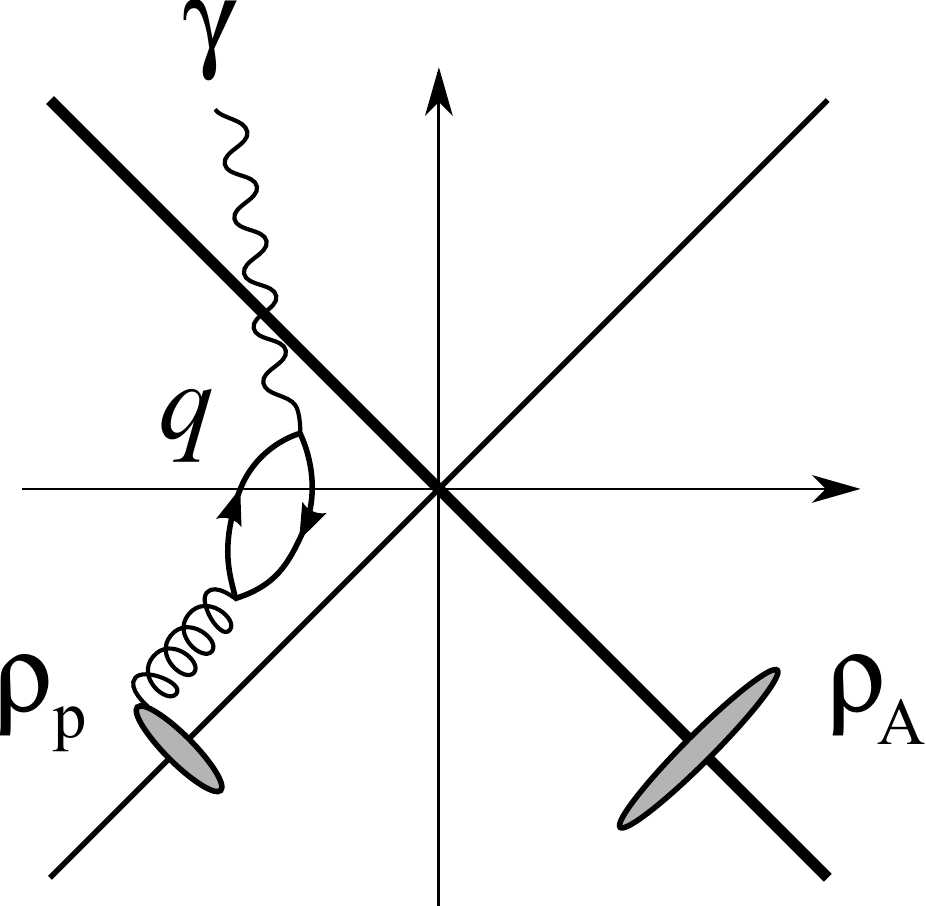}
  \end{center}
 \end{minipage}
 \hspace{6em}
 \begin{minipage}{0.3\textwidth}
  \begin{center}
   {\large (b)}
   \includegraphics[width=\textwidth]{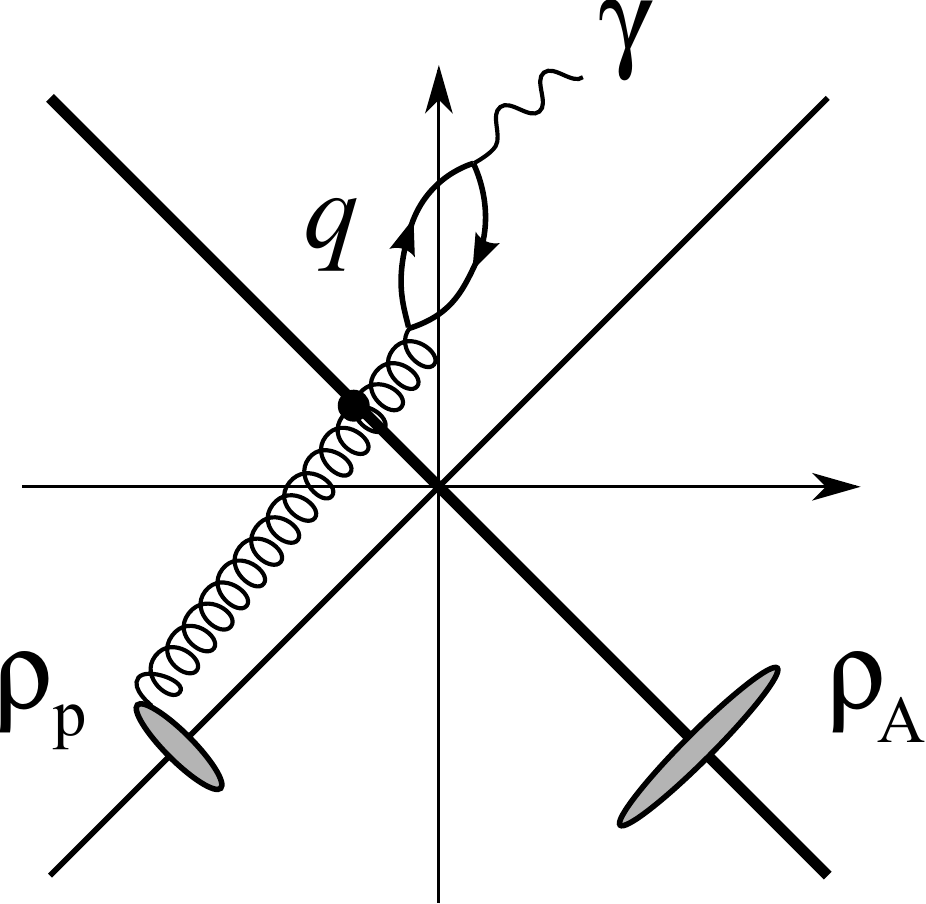}
  \end{center}
 \end{minipage}
 \vspace{1em}\\
 \begin{minipage}{0.3\textwidth}
  \begin{center}
   {\large (c)}
   \includegraphics[width=\textwidth]{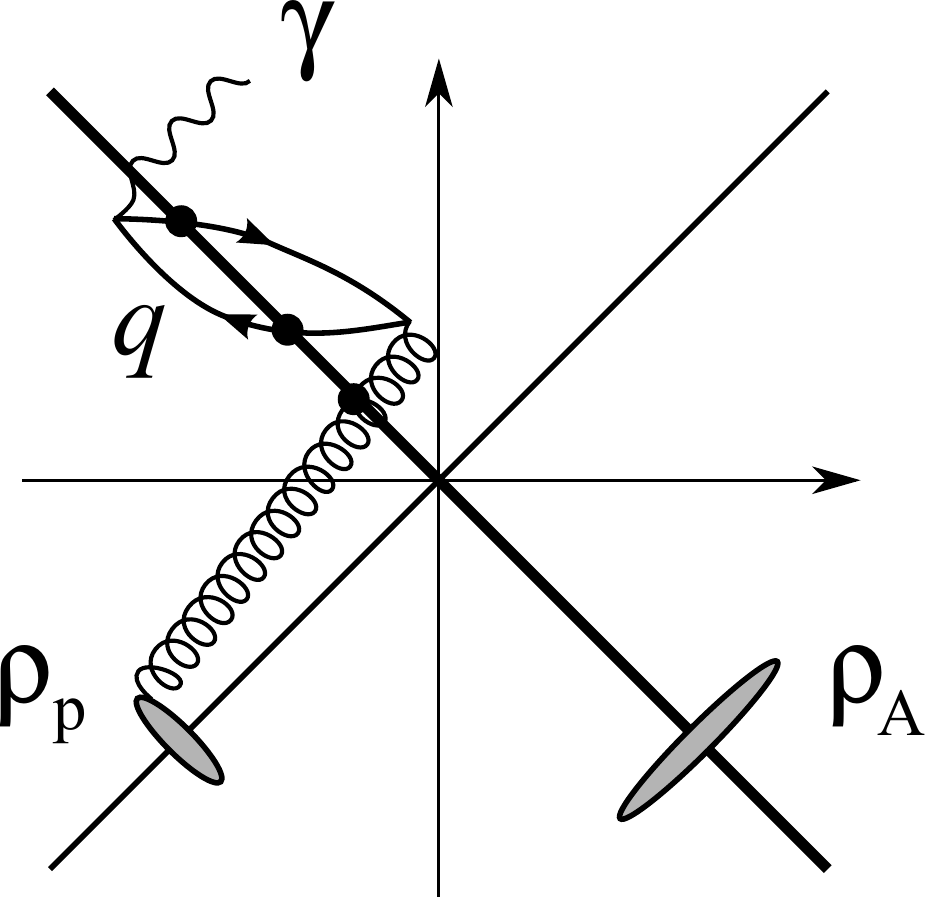}
  \end{center}
 \end{minipage}
 \hspace{6em}
 \begin{minipage}{0.3\textwidth}
  \begin{center}
   {\large (d)}
   \includegraphics[width=\textwidth]{NLO.pdf}
  \end{center}
 \end{minipage}
 \end{center}
 \caption{The connected $O(\rho_p^1)$ contributions to the amplitude $\calM_\lambda(\bk)$ corresponding to the decomposition in Eq.~\eqref{eq:mspl}.}
 \label{fig:NLOALL}
\end{figure}

Next, we focus on the third (connected) $O(\rho_p^1)$ contribution.
It is convenient to decompose the amplitude using Eq.~\eqref{eq:qprop}
for $S_{(0)}(x,y)$, which leads to
\be
\begin{split}
 \calM_\lambda(\bk) &= e g\int d^4 x\, d^4 y\, e^{ik\cdot x}\, \Tr
  \bigl[\slashchar{\epsilon}_\lambda(\bk) S_{(0<<)}(x,y)
  \slashchar{\calA}_{(1<)}(y) S_{(0<<)}(y,x)\bigr]\\
 & +e g \int d^4 x\, d^4 y\, e^{ik\cdot x}\,\Tr\bigl[
  \slashchar{\epsilon}_\lambda(\bk) S_{(0>>)}(x,y)
  \slashchar{\calA}_{(1>)}(y) S_{(0>>)}(y,x)\bigr]\\
 & +e g \int d^4 x\, d^4 y\, e^{ik\cdot x}\, \Tr\bigl[
  \slashchar{\epsilon}_\lambda(\bk) S_{(0<>)}(x,y)
  \slashchar{\calA}_{(1>)}(y) S_{(0><)}(y,x)\bigr]\\
 & +e g \int d^4 x\, d^4 y\, e^{ik\cdot x}\, \Tr\bigl[
  \slashchar{\epsilon}_\lambda(\bk) S_{(0><)}(x,y)
  \slashchar{\calA}_{(1<)}(y) S_{(0<>)}(y,x)\bigr]\;.
\end{split}
\label{eq:mspl}
\ee
We depict the graphical representation of these individual
contributions in Fig.~\ref{fig:NLOALL}.  In the first (a) term the
quark-antiquark pair is created by the gluon from $\rho_p$ and
annihilated to a photon without crossing $\rho_A$.  In the second (b)
and the third (c) terms the gluon crosses $\rho_A$ and then produces a
quark-antiquark pair.  The created pair subsequently annihilates to a
photon in the second term, but in the third term the pair first
crosses $\rho_A$ back before annihilation.  In the fourth (d) term the
pair is created prior to the interaction with $\rho_A$ and after the
created pair crosses $\rho_A$ it annihilates to a photon.

We now explain in details how the first three contributions in
Eq.~\eqref{eq:mspl} vanish.  Using Eq.~\eqref{eq:spliprop} we can
write the contribution (a) as
\be
 \int d^4 x\, d^4 y\, \theta(-x^+)\theta(-y^+) \, e^{ik\cdot x}
 \, \Tr\bigl[ \slashchar{\epsilon}_\lambda(\bk)
  S_F(x-y)\slashchar{\calA}_{(1<)}(y) S_F(y-x)\bigr]\;.
\ee
In the region $x^+, y^+<0$ the gluon field cannot develop a singlet
component.  The first contribution is then zero simply because of the
color trace $\Tr(T_F^a)=0$.  For the second component (b) with
$x^+, y^+>0$, because the CGC field is localized in $x^+$ as in
Eq.~\eqref{eq:A0}, the quark loop part takes an identical structure as
that of (a), and the quark loop picks up the color trace
$\Tr(T_F^a)=0$.  It is quite straightforward to confirm that the third
component (c) is vanishing from the complex pole structures.  We can
intuitively understand this from the directions of particle and
anti-particle flows in the light-cone coordinates as mentioned around
Eq.~\eqref{eq:propagation}:  a positive energy should always flow
from smaller to larger $x^+$.  We also make a remark that (c) is
clearly zero just because the quark-pair creation and annihilation
points are not causally connected.

Thus, the only remaining contribution appears from the fourth term in
Fig.~\ref{fig:NLOALL}~(d).  Transforming the integrand in momentum
space we can write the total $O(\rho_p^1)$ amplitude as follows:
\begin{align}
 \calM_\lambda(\bk) & = e g \int \frac{d^4 p}{(2\pi)^4}
  \frac{d^4 p'}{(2\pi)^4}\frac{d^4 q}{(2\pi)^4} (2\pi) \delta(p'^+)
  (2\pi) \delta(p^+ + p'^+ - k^+) \notag\\
  &\quad\times \theta(q^+)\theta(k^+-q^+)\text{Tr}_c\bigl[
    U(\bpt')\calA_{(1<)}^\mu(p)U^\dag(\bpt+\bpt'-\bkt)\bigr] \notag\\
  &\quad\times \text{Tr}_D\bigl[
   \slashchar{\epsilon}_\lambda(\bk) S_F(q)\slashchar{n}
   S_F(q-p')\gamma_\mu S_F(q-p-p') \slashchar{n}S_F(q-k)\bigr]\;.
\label{eq:mgl}
\end{align}
In the physical language, $q$ is the momentum running over the quark
loop, $p$ is the momentum carried by the gluon field attached with
$\rho_p$, $p'$ and $p+p'-k$ are the momenta inserted by the
interaction with the CGC gluon field from $\rho_A$.  We note that the
$\theta$-functions are to keep the correct energy (longitudinal
momentum) flows.  Because the CGC fields convey only the transverse
momenta, the integrals with respect to $p^+$ and $p'^+$ are trivially
constrained by the $\delta$-functions.  The integrals over $p^-$,
$p'^-$ and $q^-$ are computed as we explain below.  Taking into
account the gluon field $\mathcal{A}_{(1<)}^\mu(p)$ as given in
Eq.~\eqref{eq:gl1}, we see that the integral over $p^-$ has two
singularities; one above and the other below the real axis.  We shall
perform the $p^-$-integration by picking up the singularity in
$S_F(q-p-p')$ at
\be
 p^- =  q^- - p'^- - \frac{\omega_{q-p-p'}^2}{2(q^+-k^+)}-i\epsilon\;.
\label{eq:sip}
\ee
Here we introduced a notation for the transverse energy as
$\omega_p^2\equiv p_\perp^2+m^2$.  Next, we calculate the integration
over $p'^-$.  Also in this case we find two singularities above and
below the real axis.  We pick up the singularity in $S_F(q-p')$ at
\be
 p'^- = q^- - \frac{\omega_{q-p'}^2}{2 q^+} + i\epsilon\;.
\label{eq:sipp}
\ee
The gluon fields $\calA_{(1<)}^\mu$ in Eq.~\eqref{eq:gl1} have
non-vanishing transverse components.  Plugging Eqs.~\eqref{eq:sip} and
\eqref{eq:sipp} into Eq.~\eqref{eq:gl1} we have $\calA^i_{(1<)}(p)$ in
the following form,
\be
 \calA^i_{(1<)}(p) = ig \frac{1}{(k^+ +i\epsilon)\Bigl(
   \frac{\omega_{q-p'}^2}{2q^+}-\frac{\omega_{q-p-p'}^2}{2(q^+ - k^+)}
   -i\epsilon\Bigr)}\, \frac{p^i}{p_\perp^2} \,\rho_p(\bpt)\;,
\ee
where the $q^-$ dependence has canceled out.  Therefore, for the
$q^-$-integration there are two remaining singularities; one above the
real axis from $S_F(q-k)$ and the other below the real axis from
$S_F(q)$.  We choose to pick up the singularity in $S_F(q)$ at
\be
 q^- = \frac{\omega^2_q}{2q^+}-i\epsilon\;.
\ee
Now, we still have the $q^+$-integration and three transverse
integrations with respect to $\bpt$, $\bpt'$, and $\bqt$.  The
amplitude has two non-trivial denominators coming from
$\calA^i_{(1<)}(p)$ and $S_F(q-k)$.  We can further simplify the
singular denominator as
\be
\begin{split}
  &\biggl[ \frac{\omega_{q-p'}^2}{2q^+}
    -\frac{\omega_{q-p-p'}^2}{2(q^+ - k^+)}\biggr]
  \biggl[2(q^+-k^+)\biggl(\frac{\omega_q^2}{2q^+} - k^-\biggr)
    - \omega_{q-k}^2\biggr]\\
  &= \frac{k^{+2}}{2q^{+2}(q^+ - k^+)}\bigl[ (x\bkt-\bqt)^2+m^2\bigr]
  \bigl[ (x\bpt-\bqt+\bpt')^2+m_p^2(x)\bigr]\;,
\end{split}
\label{eq:denom}
\ee
where $x\equiv q^+/k^+$.  Later we will use $x$ as an integration
variable instead of $q^+$.  In the above we defined
$m_p^2(x) \equiv m^2 + b(x)p_\perp^2$ with $b(x) \equiv x(1-x)$.

Let us turn to the calculation of the numerator.  We have computed the
Dirac trace with the help of FeynCalc~\cite{Mertig:1991}.  Using the
explicit form of the polarization vector,
\be
 \epsilon_{\lambda\mu}(\bk) = g_{\lambda\mu}-\frac{n_\lambda k_\mu
  + n_\mu k_\lambda}{n\cdot k}
\ee 
with the physical polarizations $\lambda=1,2$, we find that the Dirac
trace eventually leads to
\be
\begin{split}
  &\Tr_D\bigl[\slashchar{\epsilon}_\lambda(\bk) (\slashchar{q}+m)
    \slashchar{n}(\slashchar{q}-\slashchar{p}'+m)
    \bpt\cdot \boldsymbol{\gamma}_\perp
    (\slashchar{q}-\slashchar{p}-\slashchar{p}' + m)
    \slashchar{n}(\slashchar{q}-\slashchar{k}+m) \bigr]\\
  &= 8k^{+2}\bigl[ R_1(\bpt,\bpt',\bqt)(xk_\lambda-q_\lambda)\\
  &\qquad\qquad +R_2(\bkt,\bpt,\bqt)
  (x p_\lambda - q_\lambda + p'_\lambda) + R_3(\bkt,\bpt,\bqt) p_\lambda\bigr]~,
\end{split}
\label{eq:dirt1}
\ee
where we defined,
\begin{equation}
 \begin{split}
   R_1(\bpt,\bpt',\bqt) &\equiv a(x)b(x)\bpt^2
   - [1-4b(x)](x\bpt -\bqt+\bpt')\cdot \bpt\;, \\
   R_2(\bkt,\bpt,\bqt) &\equiv (x\bkt-\bqt)\cdot \bpt\;, \\
   R_3(\bkt,\bpt,\bqt) &\equiv (x\bkt-\bqt)\cdot \bpt - m^2\;,
 \end{split}
\end{equation}
and $a(x) \equiv x-\frac{1}{2}$, and see below Eq.~\eqref{eq:denom}
for the definition of $b(x)$.  Finally, we put all the terms together
to rewrite the amplitude as
\begin{align}
 \calM_\lambda(\bk) &= \frac{e g^2}{\pi}\int_0^1 dx\int
 \frac{d^2 \bpt}{(2\pi)^2}\frac{d^2 \bpt'}{(2\pi)^2}\frac{d^2 \bqt}{(2\pi)^2}
 \,\frac{1}{p_\perp^2} \, \frac{1}{(x \bkt-\bqt)^2+m^2} \notag\\
 &\qquad \times\frac{1}{(x\bpt-\bqt+\bpt')^2+m_p^2(x)}
  \Bigl\{R_1(\bpt,\bpt',\bqt)(xk_\lambda-q_\lambda) \notag\\
 &\quad +R_2(\bkt,\bpt,\bqt)
  (x p_\lambda-q_\lambda + p_\lambda') + R_3(\bkt,\bpt,\bqt)
  p_\lambda\Bigr\}
  \notag\\
 &\qquad \times \text{Tr}_c \bigl[ U(\bpt')\,
 \rho_p(\bpt)\,U^\dag(\bpt+\bpt'-\bkt)\bigr] \;.
\label{eq:mmom}
\end{align}
We now make a few general comments about the above amplitude.  First,
decomposing the amplitude as
$\calM_\lambda(\bk)=\epsilon^\mu_\lambda(\bk) \mathcal{T}_\mu(\bk)$,
we have explicitly checked that the Ward identity,
$k_\mu \mathcal{T}^\mu(\bk) = 0$, holds.  The relevant steps of this
calculation are summarized in \ref{sec:appa}.  Second, we note that
the numerator depends on a combination of $x\bkt-\bqt$ and the same
dependence is found in the denominator.  This leads to a consequence
that any collinear singularity in the $\bqt$-integration even at $m=0$
does not appear due to the cancellation by the numerator.

At this point it is useful to replace the integration variable as
$\bpt' \to \blt \equiv \bpt' - \bqt +\frac{1}{2}\bpt$.  We transform
the Wilson lines back in position space for convenience.  Shifting the
integration variables as
$x\bkt - \bqt \to \bqt$ and $a(x)\bpt + \blt \to \blt$, we can
evaluate the transverse momentum integrals with the help of
\be
\begin{split}
  \int\frac{d^2 q_\perp}{(2\pi)^2}\,
   \frac{e^{i\bqt\cdot\bxt}}{q_\perp^2 +m^2}
   &= \frac{1}{2\pi}K_0(x_\perp m)\;,\\
  \int\frac{d^2 q_\perp}{(2\pi)^2}\,
   \frac{e^{i\bqt\cdot\bxt}}{q_\perp^2 +m^2} \, \bqt\cdot \byt
   &= \frac{i}{2\pi}\, \hat{\boldsymbol{x}}_\perp\cdot \byt \,
    m K_1(x_\perp m)\;,
\end{split}
\ee
where $\hat{\boldsymbol{x}}_\perp \equiv \bx/x_\perp$ and
$K_{0,1}(x)$ are the modified Bessel functions of the zeroth and the
first order, respectively.  After all, the final expression for the
amplitude that we will use for the computation of the photon
production rate is
\begin{align}
 \calM_\lambda(\bk) &= \frac{e g^2}{4\pi^3}\int_0^1 dx\int
  d^2 \bu\, d^2 \bv \int \frac{d^2 \bpt}{(2\pi)^2}\,
  e^{-i(\bkt-\bpt) \cdot [\bv+a(x)\bu]}\notag\\
 &\quad\times \frac{\rho_p^a(\bpt)}{p_\perp^2} \, \Tr_c \biggl[
  U\Bigl(\bv+\frac{\bu}{2}\Bigr) T_F^a U^\dag\Bigl(\bv-\frac{\bu}{2}
  \Bigr)\biggr] \notag\\
  &\quad\times \bigl[\hat{u}_\lambda p_\perp \Psi_1(\bpt,\bu,x)
    + p_\lambda \Psi_2(\bpt,\bu,x) \bigr]\;,
\label{eq:ampl2}
\end{align}
where
\begin{align}
  \Psi_1(\bpt,\bu,x) &\equiv -4i a(x)b(x) p_\perp
   K_0(m_p(x) u_\perp) m K_1(m u_\perp) \notag\\
  &\quad +4b(x)\hbp\cdot\hbu
   m_p(x) K_1(m_p(x) u_\perp) m K_1(m u_\perp)\;,\\
  \Psi_2(\bpt,\bu,x) &\equiv m K_1(m u_\perp) m_p(x)
   K_1(m_p(x) u_\perp) \notag\\
  &\quad +m^2 K_0(m u_\perp) K_0(m_p(x) u_\perp)\;.
\end{align}
These functions $\Psi_1$ and $\Psi_2$, have the following symmetry
properties:
\be
\begin{split}
 \Psi_i(\bpt,\bu,x) &= \Psi_i^*(\bpt,-\bu,x);,\\
 \Psi_i(\bpt,\bu,x) &= \Psi_i^*(\bpt,\bu,1-x)\;,\\
 \Psi_i(\bpt,\bu,x) &= (-1)^i\Psi_i(\bpt,-\bu,1-x)\;.
\end{split}
\label{eq:psisym}
\ee

It should be noted that the Wilson lines take care of resummation over
multiple gluon scatterings and we can recover the naive diagrammatic
perturbation theory by expanding the Wilson lines in the number of the
gluon fields.  The contribution with a single gluon (i.e.\ no gluon
from the Wilson lines) vanishes because of the color trace.  The
contribution with two gluons vanishes because of charge conjugation or
Furry's theorem.  The lowest non-vanishing diagram should involve
three gluons, namely, one from $\rho_p$ and two from $\rho_A$, and an
emitted photon.  It is important to realize that this contribution is
UV finite due to gauge invariance.  That is, if a three-gluon and
one-photon operator came from a UV divergent loop, it would have to
correspond to a dimension four operator, but there is no such gauge
invariant operator with dimension four involving three gluons and one
photon.  Because higher-order contributions are more UV suppressed,
our resummed result in Eq.~\eqref{eq:ampl2} is completely UV finite.

\section{Photon production rate}
\label{sec:rat}

Squaring the amplitude we obtain the probability density for the
emission of a single photon.  This expression explicitly depends on
the color sources, $\rho_p$ and $\rho_A$, and we should take a color
average over the color sources in the proton and in the nucleus.  To
get the minimum biased photon production rate we integrate over the
impact parameters $\bbt$.  In total, the photon production rate, that
is, the number of photons produced per unit $\bkt$ and per unit
rapidity $y \equiv\frac{1}{2}\ln(k^+/k^-)$ is given as
\be
 dN = \frac{d^3 \bk}{(2\pi)^3 (2k^0)}
  \int d^2 \bbt\, \bigl\langle\calM_\lambda(\bk)
  \calM_\lambda^\ast(\bk)\bigr\rangle
  = \frac{d^2 \bkt dy}{16 \pi^3}
  \int d^2 \bbt\, \bigl\langle\calM_\lambda(\bk)
  \calM_\lambda^\ast(\bk)\bigr\rangle\;,
\label{eq:rate1}
\ee
where a summation over $\lambda$ is implied.  With
$\langle\cdots\rangle$ we denoted taking the color average, that is
defined for a general operator $\calO[\rho_p,\rho_A]$ as
\be
 \bigl\langle \calO[\rho_p,\rho_A] \bigr\rangle
  = \int [d\rho_p][d\rho_A] W_p[x_p;\rho_p]W_A[x_A;\rho_A]
  \calO[\rho_p,\rho_A]\;,
\ee
where the functionals, $W_{p,A}[x_{p,A};\rho_{p,A}]$, incorporate the
small-$x$ evolution of the proton and the nucleus wave-functions.  The
non-linear evolution of $W_{p,A}[x_{p,A};\rho_{p,A}]$ is governed by
the Balitsky-Jalilian--Marian-Iancu-McLerran-Leonidov-Kovner
(B-JIMWLK) equation~\cite{Balitsky:1995ub,JalilianMarian:1997jx}, and
we will later adopt a Gaussian approximated solution for numerical
calculations.

\subsection{Squaring the amplitude}

For variables in the complex conjugated amplitude, we will use primes
on the coordinates $\bu$, $\bv$, the gluon momentum $\bpt$, the
momentum fraction $x$, and the color index $a$, which all characterize
the amplitude~\eqref{eq:ampl2}.  For the color average over $\rho_p$
we follow the notation of Ref.~\cite{Blaizot:2004wv} to define the
unintegrated gluon distribution function as
\be
g^2 \bigl\langle \rho_p^a(\bpt) \rho_p^{\dag a'}(\bpt)\bigr\rangle
 \equiv \frac{\delta^{aa'}}{\pi (N_c^2-1)} p_\perp^2\, \varphi_p(p_\perp)\;,
\ee
where the color average is taken for the proton.  Transverse momentum
integral over $\varphi_p(p_\perp)$ will be proportional $n_g$ that we
discussed in Sec.~\ref{sec:processes} (see Eq.~\eqref{eq:ugd} for a
precise relation).  The photon production rate is also proportional to
the Wilson line product, and we define,
\be
  S(\by,\bz,\by',\bz')
  \equiv \frac{1}{N_c}\Bigl\langle \Tr_c \bigl[
  U(\by) T_F^a U^\dag(\bz)\bigr]\Tr_c \bigl[ U(\bz') T_F^a
  U^\dag(\by')\bigr]\Bigr\rangle \;,
\label{eq:trc2}
\ee
where the color average is taken for the nucleus.  We have the
original coordinate variables through $\by = \bv - \bu/2$,
$\bz = \bv + \bu/2$ and $\by' = \bv' - \bu'/2$, and
$\bz' = \bv' + \bu'/2$.  Next, we shift the coordinates of the proton
position as $\bv \to \bv - \bbt$ and $\bv' \to \bv' - \bbt$.  In this
way we moved the origin of the coordinate system from the proton to
the nucleus.  In the infinite limit of nucleus transverse size the
correlator~\eqref{eq:trc2} is invariant under translation in the
transverse plane and therefore independent of $\bbt$.  The integration
over the impact parameter thereby results in a factor,
\be
 \int d^2 \bbt\, e^{-i(\bpt-\bpt')\cdot \bbt}
 = (2\pi)^2 \delta^{(2)}(\bpt - \bpt')\;.
\label{eq:imp}
\ee
As a further consequence from translational invariance, the
correlator~\eqref{eq:trc2} is independent of the overall center of
mass coordinate, $\bv + \bv'$, so that the integration over $\bv$ and
$\bv'$ is reduced to
\be
 \int d^2 \bv d^2 \bv' = \pi R_A^2 \int d^2 \bw\;,
\ee
where $\pi R_A^2$ is the transverse area of the nuclei and
$\bw \equiv \bv - \bv'$.

For a given polarization $\lambda$, the amplitude~\eqref{eq:ampl2} is
a linear combination of $\hat{u}_\lambda$ and $p_\lambda$.  Squaring
the amplitude and summing over $\lambda$ results in four terms
proportional to $\Psi_1\Psi'^{\ast}_1 \, \hbu\!\!\cdot\hbu'$,
$\Psi_2\Psi'^{\ast}_2$,  $\Psi_1\Psi'^{\ast}_2 \, \hbu\!\!\cdot\hbp$,
and $\Psi_2\Psi'^{\ast}_1 \, \hbp\!\!\cdot\hbu'$, respectively.  Here
we used an abbreviated notation
$\Psi_{1,2} \equiv \Psi_{1,2}(\bpt,\bu,x)$ and
$\Psi'_{1,2} \equiv \Psi_{1,2}(\bpt',\bu',x')$.  The last two terms
are shown to be equal by exchanging the original and the primed
coordinates and momentum fractions, that is,
$\bu \leftrightarrow \bu'$, $x \leftrightarrow x'$, followed by a
reflection $\bu \to -\bu$ and $\bu' \to -\bu'$.  Such a transformation
leaves the exponential factor
$e^{-i(\bkt - \bpt)\cdot [\bv-\bv' + a(x)\bu-a(x')\bu']}$ as well as
the Wilson line product~\eqref{eq:trc2} intact.  Due to the
relations~\eqref{eq:psisym}, we can write
$\Psi_1\Psi'^{*}_2 \, \hbu\cdot\hbp = \Psi_2\Psi'^{*}_1 \,
\hbp\cdot\hbu'$, under the integral.

Finally, the expression for the photon production rate takes the form
of
\begin{equation}
 \begin{split}
  &\frac{1}{\pi R_A^2}\frac{dN}{d^2 \bkt dy}
   = \frac{\alpha\, \alpha_s}{16\pi^8}\frac{N_c}{N_c^2-1}
   \int_0^1 dx\,dx'\int d^2 \bu\, d^2 \bu'\, d^2 \bw\, e^{-i\bkt\cdot\br}\\
  &\quad\times S\Bigl(\bu-\frac{\bv}{2},\bu+\frac{\bv}{2},
   \bu'-\frac{\bv'}{2},\bu'+\frac{\bv'}{2}\Bigr)\\
  &\quad\times\int \frac{d^2 \bpt}{(2\pi)^2}
   \frac{p_\perp^4\varphi_p(p_\perp)}{(p_\perp^2 + \LQCD^2)^2}\,
   e^{i\bpt\cdot\br} \bigl( \hbu\!\!\cdot\hbu'\Psi_1\Psi'^\ast_1
   + \Psi_2 \Psi'^\ast_2 + 2\hbu\!\!\cdot\hbp\Psi_1\Psi'^\ast_2\bigr)\;,
 \end{split}
\label{eq:ratf}
\end{equation}
where we regulate the infrared divergence from massless gluons by an
infrared cutoff $\LQCD$ and we introduced a notation,
$\br \equiv \bw+a(x)\bu - a(x')\bu'$.  We see that Eq.~\eqref{eq:ratf}
is $\calO(\alpha\,\alpha_s\,n_g)$ and this expression represents one
of the main results in this work.

\subsection{Taking the color average}

Applying the Fierz transformation to the correlator~\eqref{eq:trc2} we
find,
\be
\begin{split}
 S(\by,\bz,\by',\bz') &= \frac{1}{2}\Bigl[ Q(\by,\bz,\by',\bz')\\
 &\quad -\frac{1}{N_c^2}\big\langle \Tr_c [U(\by) U^\dag(\bz)]
  \Tr_c [U(\by') U^\dag(\bz')]\big\rangle\Bigr]\;,
\end{split}
\label{eq:sinel}
\ee
where
\be
 Q(\by,\bz,\by',\bz') \equiv \frac{1}{N_c}\bigl\langle \Tr_c
  [U(\by) U^\dag(\bz')U(\by') U^\dag(\bz)]\bigr\rangle\;,
\ee
is the color average over the quadrupole operator.  The second term in
the large $N_c$ limit and for large nuclei becomes the product of the
dipole operators as
\be
 \frac{1}{N_c^2}\bigl\langle\Tr_c [U(\by) U^\dag(\bz)]
  \Tr_c [U(\by') U^\dag(\bz')] \bigr\rangle \to
 D(\by,\bz)\,D(\bz',\by')\;,
\ee
and the color average over the dipole operator is given as
\be
 D(\by,\bz) \equiv \frac{1}{N_c}\bigl\langle \Tr_c
 [U(\by) U^\dag(\bz)] \bigr\rangle\;.
\ee
The expression~\eqref{eq:sinel} coincides with the so-called
``inelastic quadrupole''~\cite{Dominguez:2011wm} (see also
Refs.~\cite{JalilianMarian:2004da,Kang:2013hta} for phenomenological
applications to gluon-gluon and quarkonium production, respectively.

The general frameworks for the small-$x$ evolution of the dipole and
the quadrupole are incorporated in the B-JIMWLK
equations~\cite{Balitsky:1995ub,JalilianMarian:1997jx}.  The dipole
evolution is closed in the large $N_c$ limit where it is known as the
Balitsky-Kovchegov (BK)
equation~\cite{Balitsky:1995ub,Kovchegov:1996ty,Kovchegov:1999ua}.
Concerning the phenomenological applications, the running coupling BK
(rcBK) equation~\cite{Albacete:2007yr} is widely used.  However, for
the quadrupole evolution such a simplification has not been found.  So
far, it has been considered only for some very specific
configurations~\cite{Iancu:2011nj}.  Eventually, one would want to
constrain the quadrupole evolution by using experimental data as it is
done for the rcBK evolution by the DIS data from HERA.

In this work, as a preliminary for going into such quantitative
studies, we will make a Gaussian approximation for the color
distribution over the nuclei $W_A[x_A;\rho_A]$, which defines the
McLerran-Venugopalan (MV) model~\cite{McLerran:1993ni} as
\be
 \bigl\langle \rho_A^a(\bxt) \rho_A^b(\byt) \bigr\rangle
 = g^2 \delta^{ab}\mu_A^2 \delta^{(2)}(\bxt-\byt)\;,
\ee
where the parameter $\mu_A$ is related to the saturation momentum
$Q_s$.  Here we shall employ a simple definition of $Q_s$ by
\begin{equation}
  Q_s^2 \equiv \frac{N_c^2-1}{4N_c} g^4 \mu_A^2 \;.
\label{eq:qsat}
\end{equation}
The MV model for the nuclei takes into account the multiple scattering
effect, and typically, the MV model is considered to work up to a
moderate value of $x_A\sim 10^{-2}$.  To reach a region with far
smaller $x_A$ we should consider the MV model as an initial condition
for the evolution equations.
 
Using the standard techniques of the MV
model~\cite{Blaizot:2004wv,Gelis:2001da,Fukushima:2007dy} we have
found it most convenient to directly calculate
$S(\by,\bz,\by',\bz')$ in position space.  The calculation steps are
collected in \ref{sec:appc} leading to the following result:
\begin{equation}
 \begin{split}
 S(\by,\bz,\by',\bz') &= \frac{1}{N_c}
  \frac{(N_c^2-1) (\beta-\alpha)}{\sqrt{N_c^2 (\alpha-\gamma)^2
  -4(\alpha-\beta)(\beta-\gamma)}}\\
  &\quad\times \exp\biggl[-\frac{2\beta+(N_c^2-2)(\alpha+\gamma)}
    {N_c^2-1}\biggr]\\
  &\quad\times \sinh\biggl[\frac{N_c}{N_c^2-1}
 \sqrt{N_c^2 (\alpha-\gamma)^2-4(\alpha-\beta)(\beta-\gamma)}\biggr]\;.
\end{split}
\label{eq:stot}
\end{equation}
We note that the above result does not rely on the large-$N_c$
limit.  Here we defined functions, $\alpha$, $\beta$, and $\gamma$ as
\begin{equation}
 \begin{split}
 & 2\alpha(\by,\bz,\by',\bz')
  \equiv B_2(|\by - \bz'|) + B_2(|\bz - \by'|) \;,\\
 & 2\beta(\by,\bz,\by',\bz')
  \equiv B_2(|\by - \by'|) + B_2(|\bz - \bz'|) \;,\\
 & 2\gamma(\by,\bz,\by',\bz')
  \equiv B_2(|\by - \bz|) + B_2(|\by' - \bz'|)\;.
 \end{split}
\label{eq:abg}
\end{equation}
with $B_2$ having the following explicit expression~\cite{Gelis:2001da},
\begin{equation}
\begin{split}
 B_2(x_\perp) &\equiv 2Q_s^2 \int_0^\infty \frac{k_\perp dk_\perp}{2\pi}
  \,\frac{1-J_0(k_\perp x_\perp)}{(k_\perp^2+\LQCD^2)^2}\\
 &= \frac{1}{2\pi}\frac{Q_s^2}{\LQCD^2}
 \Bigl[1 - x_\perp\LQCD K_1(x_\perp\LQCD)\Bigr] \;,
\label{eq:colmom}
\end{split}
\end{equation}
where we use $\LQCD$ as an infrared regulator again.  It would be
useful to point out that the above expression for $S$ has the
following symmetries:
\be
\begin{split}
 S(\bz,\by,\bz',\by') &= S(-\by,-\bz,-\by',-\bz')~,\\
 S(\bz,\by,\bz',\by') &=  S(\by,\bz,\by',\bz')\;.
\end{split}
\label{eq:symms}
\ee

Although we will not use it for our numerical calculations, we can
check that the large-$N_c$ limit simplifies the results into
\begin{equation}
 S(\by,\bz,\by',\bz') = \frac{\alpha-\beta}{2(\alpha-\gamma)}
 (e^{-2\alpha}-e^{-2\gamma}) + \calO(N_c^{-2})\;.
\label{eq:snc}
\end{equation}
This large-$N_c$ expression coincides with the one found in
Ref.~\cite{Dominguez:2011wm}.  Another useful check of
Eq.~\eqref{eq:stot} is that the photon rate involving two gluons
vanishes. 
The expansion
in the number of the gluon lines from the nucleus is equivalent to an
expansion in powers of $Q_s^2$.  Up to the order $\calO(Q_s^6)$ we
get,
\be
 S(\by,\bz,\by',\bz') = -(\alpha - \beta)
  + \frac{N_c^2-2}{N_c^2-1}(\alpha-\beta)\biggl(\alpha
  + \gamma +\frac{2\beta}{N_c^2-2}\biggr) + \calO(Q_s^6)\;,
\label{eq:pert}
\ee
where the first term is $\calO(Q_s^2)$ and the second is
$\calO(Q_s^4)$.  It is easy to confirm that the $\calO(Q_s^2)$ term is
odd under $\bu \to -\bu$ or $\bu' \to -\bu'$.  This symmetry is not
shared with the full expression~\eqref{eq:colmom}.  We can use this
symmetry to demonstrate that the rate vanishes to this order.  We
split the integration over $x$ as
$\int_0^1 dx = \int_0^{1/2} dx + \int_{1/2}^1 dx$ and in the second
term we transform the variable as $x\to 1-x$.  Transforming
$\bu \to -\bu$ and using the last line of Eq.~\eqref{eq:psisym} we
find $-\hbu \Psi_1(\bpt,-\bu,1-x) + \hbp \Psi_2(\bpt,-\bu,1-x)
= \hbu \Psi_1(\bpt,\bu,x) + \hbp \Psi_2(\bpt,\bu,x)$.
Since the first term in the expansion~\eqref{eq:pert} is odd in
$\bu \to -\bu$, two contributions from $\int_0^{1/2}dx$ and
$\int_{1/2}^1 dx$ cancel out and the $\calO(Q_s^2)$ order vanishes.

\section{Numerical results}
\label{sec:num}

In the application to the pair $\bar{q}q$ production it was shown in
Ref.~\cite{Blaizot:2004wv} that the quadrupole correlator
$Q(\by,\bz,\by',\bz')$ factorizes in the large-$N_c$ limit to a
product of two dipoles, as mentioned before, and thus the large-$N_c$
limit greatly reduces the numerical cost in the evaluation of this
process.  For the inelastic quadrupole $S(\by,\bz,\by',\bz')$ found
here for the annihilation photon production, there is a cancellation
in the leading order in the large-$N_c$ limit between the first term
and the second term in Eq.~\eqref{eq:sinel}.  As a consequence, the
next order in the expansion of Eq.~\eqref{eq:snc} does no longer
factorize.  A simultaneous numerical integration over the complete set
of coordinates characterizing the inelastic quadrupole is necessary.
In the numerical calculations below we use the form of the inelastic
quadrupole~\eqref{eq:stot} without any large-$N_c$ approximation.
Then, we must specify the unintegrated gluon distribution inside the
proton $\varphi_p(p_\perp)$ appearing in the rate~\eqref{eq:ratf}.
We will use the MV model for the proton to fix $\varphi_p(p_\perp)$ as
\be
 p_\perp^2\varphi_p(p_\perp) = 4\pi \, N_c n_g \, \pi R_p^2\;,
\label{eq:ugd}
\ee
where $\pi R_p^2$ is the transverse area of the proton and $n_g$ is
the transverse gluon density parameter with mass dimension 2.

\subsection{Details of the numerical procedure}

To explain the steps of the numerical calculation we introduce a
notation for convenience.  We replace the integration variable $\bw$
with $\br$ as defined by $\bw = \br -a(x)\bu + a(x')\bu'$.  The
benefit of this is that we can write the integration in the following
form,
\be
 \frac{1}{\pi R_A^2 \, \pi R_p^2 \, \alpha_s n_g}\cdot
 \frac{dN}{d^2 \bkt dy} = \frac{\alpha}{8\pi^8}
 \frac{N_c^2}{N_c^2-1}\int d^2 \br\, e^{-i\bkt\cdot\br}
 F(r_\perp)\;,
\label{eq:num1}
\ee
where we defined $F(r_\perp)$ as
\be
 F(r_\perp) \equiv \int_0^1 dx dx' \int d^2 \bu d^2\bu'\,
 f(x,x',\bu,\bu',\br)\;,
\label{eq:fhank}
\ee
with
\be
\begin{split}
 &f(x,x',\bu,\bu',\br) \equiv \tilde{S}(\bu,\bu',\br -a(x)\bu + a(x')\bu')\\
 &\times\int \frac{d^2 \bpt}{2\pi}
  \frac{p_\perp^4 \, e^{i\bpt\cdot\br}}{(p_\perp^2 + \LQCD^2)^2}
  \bigl(\hbu\!\!\cdot\hbu'\Psi_1\Psi'^\ast_1 + \Psi_2 \Psi'^\ast_2
  + 2\hbu\!\!\cdot\hbp\Psi_1\Psi'^\ast_2\bigr)\;.
\end{split}
\ee
The integration over the angle corresponding to the gluon momenta
$\bpt$ is given in \ref{sec:appb}.  In the above equation we have
re-labeled the functional dependence of the inelastic quadrupole as
\be
 S\biggl(\bu-\frac{\bv}{2},\bu+\frac{\bv}{2},\bu'-\frac{\bv'}{2},
 \bu'+\frac{\bv'}{2}\biggr) \equiv \tilde{S}(\bu,\bu',\bw)\;,
\ee
in order to emphasize its explicit dependence only on the difference
$\bw = \bv - \bv'$.  We should note that $F(r_\perp)$ does not depend
on the orientation of the vector $\br$ as the integrand in
Eq.~\eqref{eq:fhank} depends only on the relative angles between
$\bu$, $\bu'$, and $\br$.  The underlying reason is the rotational
invariance of the rate in the transverse plane of the photon
momentum.

The integration ranges over $x$, $x'$, and the angles corresponding to
$\bu$, $\bu'$ can be reduced by exploiting the discrete symmetries of
the integrand.  Due to the symmetries \eqref{eq:psisym} and
\eqref{eq:symms}, the integrand will not change under each of the
following two sets of transformations:
$\bu \to -\bu$, $\bu' \to -\bu'$, $\br \to -\br'$ or
$\bu \to -\bu$, $\bu' \to -\bu'$, $x\to 1-x$, $x'\to 1-x'$.
Performing the first transformation we have,
\be
\begin{split}
 \int_0^{2\pi} & d\phi_u \, d\phi_{u'} f(x,x',\bu,\bu',\br)\\ 
 &= 2\int_0^\pi d\phi_u \, d\phi_{u'}
 \bigl[f(x,x',\bu,\bu',\br) + f(x,x',\bu,-\bu',\br)\bigr]\;,
\end{split}
\ee
where $\phi_u$ and $\phi_{u'}$ are polar angles corresponding to $\bu$
and $\bu'$, respectively.  Using the second transformation we can
rewrite the original integration as
\be
\begin{split}
 \int_0^1 dx \, dx' &\int_0^{2\pi} d\phi_u \, d\phi_{u'}
 f(x,x',\bu,\bu',\br) = 4\int_0^{1/2} dx \, dx'
 \int_0^\pi d\phi_u \, d\phi_{u'}\\
 &\times\bigl[f(x,x',\bu,\bu',\br) + f(x,x',\bu,-\bu',\br)\\ 
 &\quad + f(x,1-x',\bu,\bu',\br) + f(x,1-x',\bu,-\bu',\br)\bigr]\;.
\end{split}
\label{eq:red}
\ee
Suppose the original integration region contains $N^4$ uniformly
distributed points, and then, owing to the reduction of the
integration region as given above, the numerical cost is reduced by
$(N/2)^4 - 4$.  This leads to a large reduction factor for $N>10$
typically.

\begin{figure}
 \begin{center}
 \includegraphics[width=0.8\textwidth]{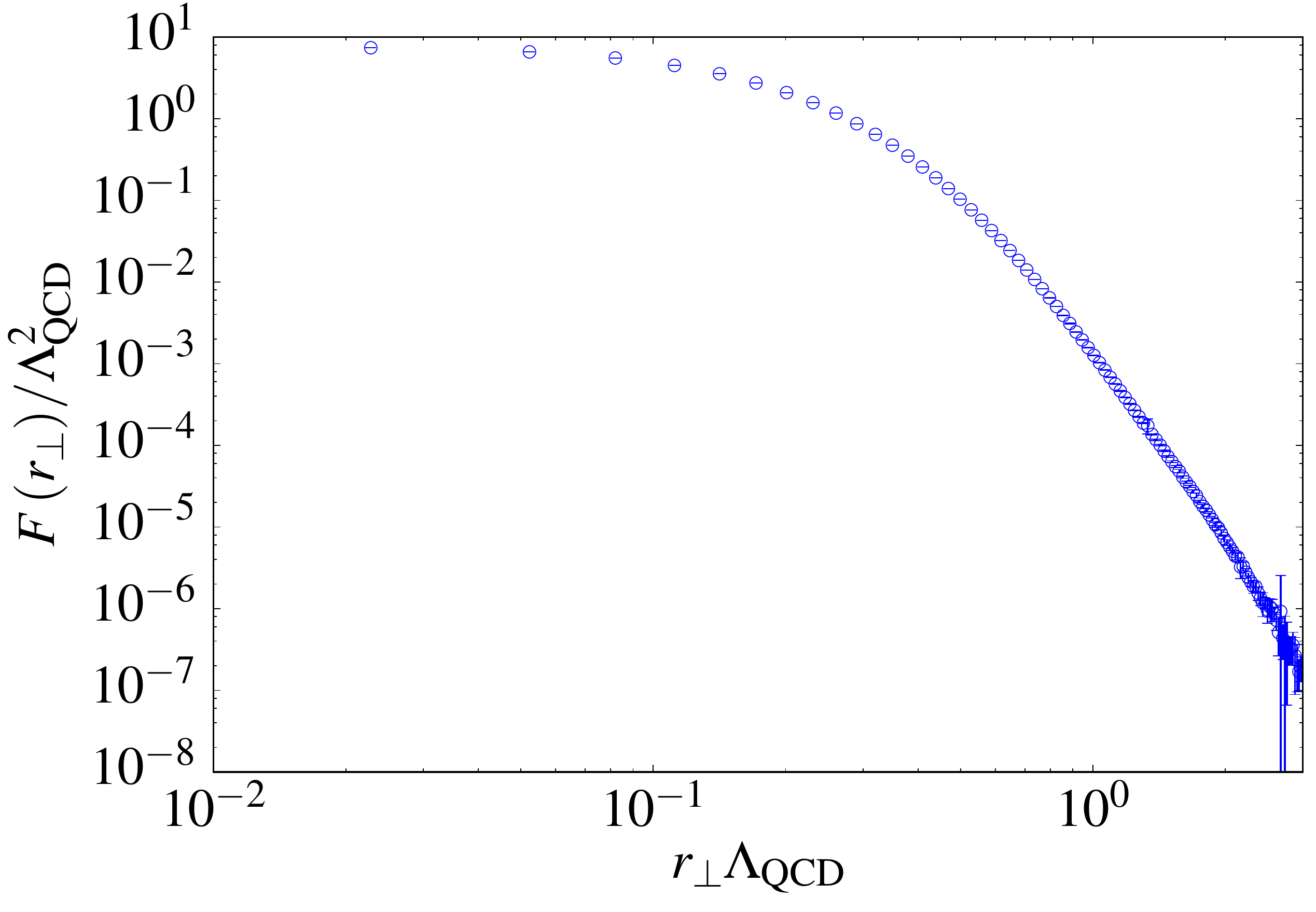}\
 \end{center}
 \caption{The numerical result for $F(r_\perp)$ defined in
   Eq.~\eqref{eq:fhank} for $Q_s = 10\LQCD$ and $m=0$ evaluated in the
   QDHT algorithm.  The error is a $3\sigma$ estimate from the MISER
   Monte Carlo algorithm.}
 \label{fig:hankel}
\end{figure}

We choose two values of the saturation momentum as
$Q_s=5\LQCD\simeq 1\;\text{GeV}$, 
$Q_s=10\LQCD\simeq 2\;\text{GeV}$ and we take the limit of vanishing
quark mass $m=0$ for numerical calculations.  We calculate
$F(r_\perp)$ by performing the 7-dimensional integration by means of
the Monte-Carlo stratified sampling MISER
algorithm~\cite{Press:1992zz}.  In our calculation we use the Python
based Scikit-Monaco package%
\footnote{\url{http://scikit-monaco.readthedocs.org/en/latest/index.html}}.
We have sampled $10^8$ integration points for each value of
$r_\perp$.  The result for $F(r_\perp)$ in the case $Q_s = 10\LQCD$ is
shown in Fig.~\ref{fig:hankel}.  

We have found that $F(r_\perp)$ rapidly decreases as we increase
$r_\perp$ which eventually can be attributed to the saturation
effect.  The Monte Carlo calculation is quite precise for small values
of $r_\perp$, while the relative error increases as $r_\perp$
increases.  For $r_\perp \LQCD \simeq 3$ the numerical value of the
function is already extremely small as
$F(3/\LQCD)/\LQCD^2 \simeq 10^{-7}$ up to $10\%$ error.  Let us
elucidate the actual numerical procedures in more details below.

The angular integration in Eq.~\eqref{eq:num1} defines the Hankel
transform for the function $F(r_\perp)$, i.e.
\be
 \int d^2\br\, e^{-i\bkt\cdot\br} F(r_\perp)
 = 2\pi \int_0^\infty r_\perp dr_\perp J_0(k_\perp r_\perp) F(r_\perp)\;,
\ee
where $J_0(x)$ is the zeroth order Bessel function.  To calculate the
above in practice, we use the numerical method known as the Quasi
Discrete Hankel Transform (QDHT)~\cite{Yu:1998}\footnote{We thank
  Francois~Gelis for suggesting the QDHT algorithm and for sharing
  with us his note on the numerical procedure.}.
The computation of the function $F(r_\perp)$ is performed on a grid
corresponding to the points prescribed by the QDHT algorithm.  In the
case $Q_s = 10\LQCD$ the maximum value on the grid is chosen as
$r^{\text{max}}_\perp \LQCD = 3$ and for the case
$Q_s = 5\LQCD$ we have taken $r^{\text{max}}_\perp \LQCD = 6$.
The minimal value of $r_\perp$ is set by fixing the number of grid
points within the QDHT algorithm.  In the calculation of $F(r_\perp)$
we used $10^2$ points.  We have tested the sensitivity to the cutoffs
imposed by the QDHT algorithm.  In particular, we confirmed that the
results up to $k_\perp \simeq 7 Q_s$ are numerically reliable.  This
is the maximum value shown in our final numerical results in
Fig.~\ref{fig:photon}.

\subsection{Discussion of the results}

\begin{figure}
 \begin{center}
 \includegraphics[width=0.8\textwidth]{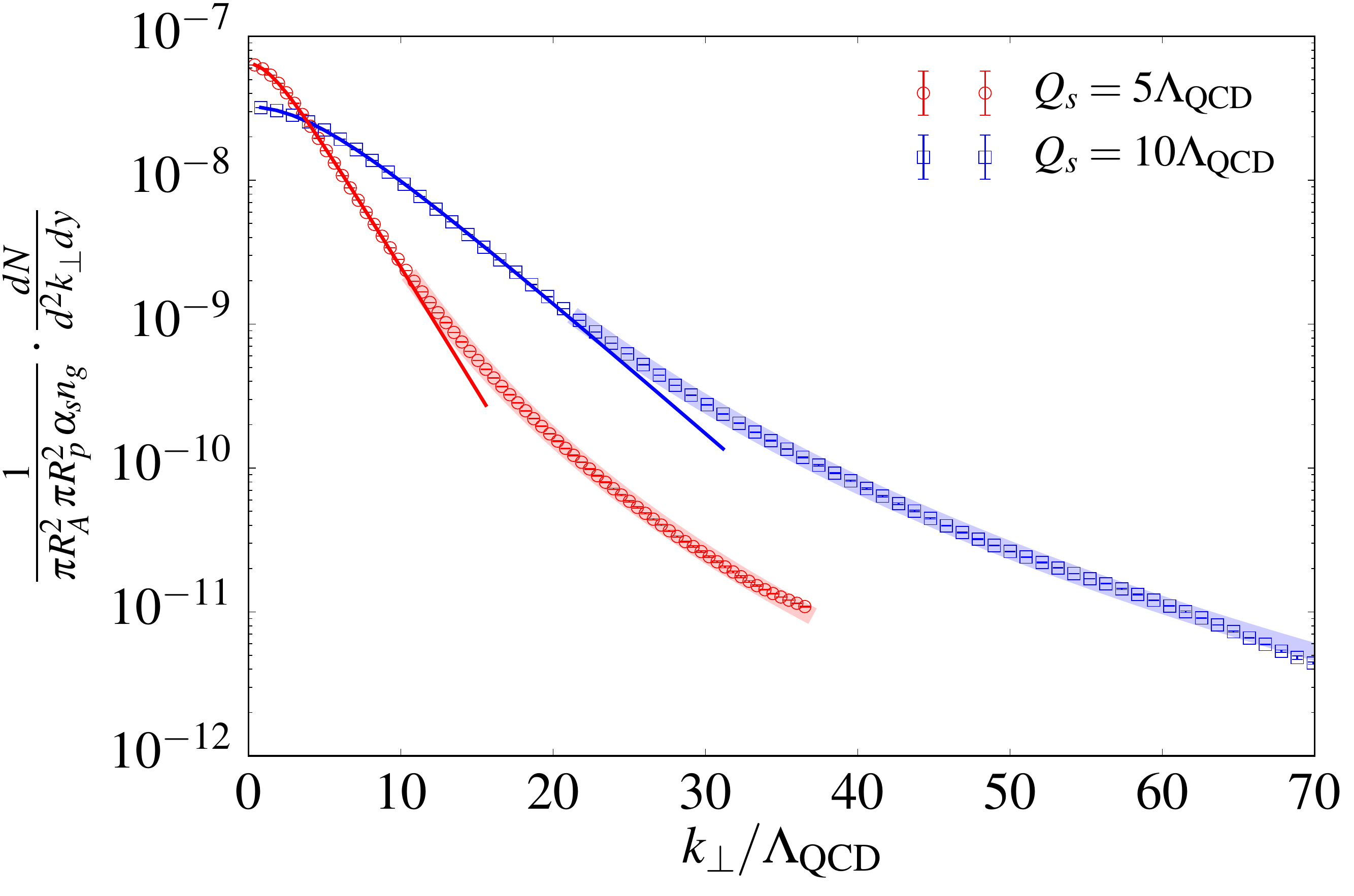}\
 \end{center}
 \caption{Photon production rate as a function of $k_\perp$ in the
   unit of $\LQCD$ in the limit of vanishing quark mass and for
   $Q_s = 5\LQCD$ and $Q_s = 10\LQCD$.  The thin lines represent the
   exponential fit, while the thick lines with light colors correspond
   to the power-law type fit.}
 \label{fig:photon}
\end{figure}

We show the numerical results for the photon spectrum in
Fig.~\ref{fig:photon} as a function of the transverse momentum
$k_\perp$.  We see that the curve slightly flattens at low momentum,
which is attributed to the saturation property.  For the results in
Fig.~\ref{fig:photon} we consider the case of a single quark flavor
with vanishing quark mass and use $\alpha=1/137$.  Although
$n_q\ll n_g\ll Q_s^2$ by definition of the semi-CGC regime of our
present interest, there is some theoretical uncertainty in precisely
determining $n_g$ of the proton.  Thanks to the simple linear
dependence on $n_g$ as a result of the expansion in terms of $\rho_p$,
we present our numerical results by scaling $n_g$ out entirely.

The data points on Fig.~\ref{fig:photon} correspond to the results
from the QDHT transform, while the solid lines represent the fit
results.  The soft part of the spectrum up to $k_\perp \sim 2Q_s$ is
very well described with a exponential fitting function,
$\exp\bigl( -\sqrt{k_\perp^2+(0.5Q_s)^2}/0.5Q_s \bigr)$.  As an
alternative, a Lorentzian-type fitting function,
$(k^2+(1.3Q_s)^2)^{-2.4}$, can work as nicely as the exponential form.
The semi-hard part for $k_\perp \gsim 2Q_s$ can be fitted by the
perturbative power-law tail as
$(\log(k_\perp/Q_s))^{1.5}/k_\perp^{5.6}$.  In Fig.~\ref{fig:photon}
we show the exponential fit by the thin lines and the power-law fit by
the thick lines with light colors.  In Ref.~\cite{McLerran:2014hza}
the Glasma photons would yield a thermal-like spectrum in the $AA$
collisions, and our calculations partially support this for
$k_\perp \lsim 2Q_s$, though a Lorentzian shape can be another
choice.

\begin{figure}
 \begin{center}
 \includegraphics[width=0.8\textwidth]{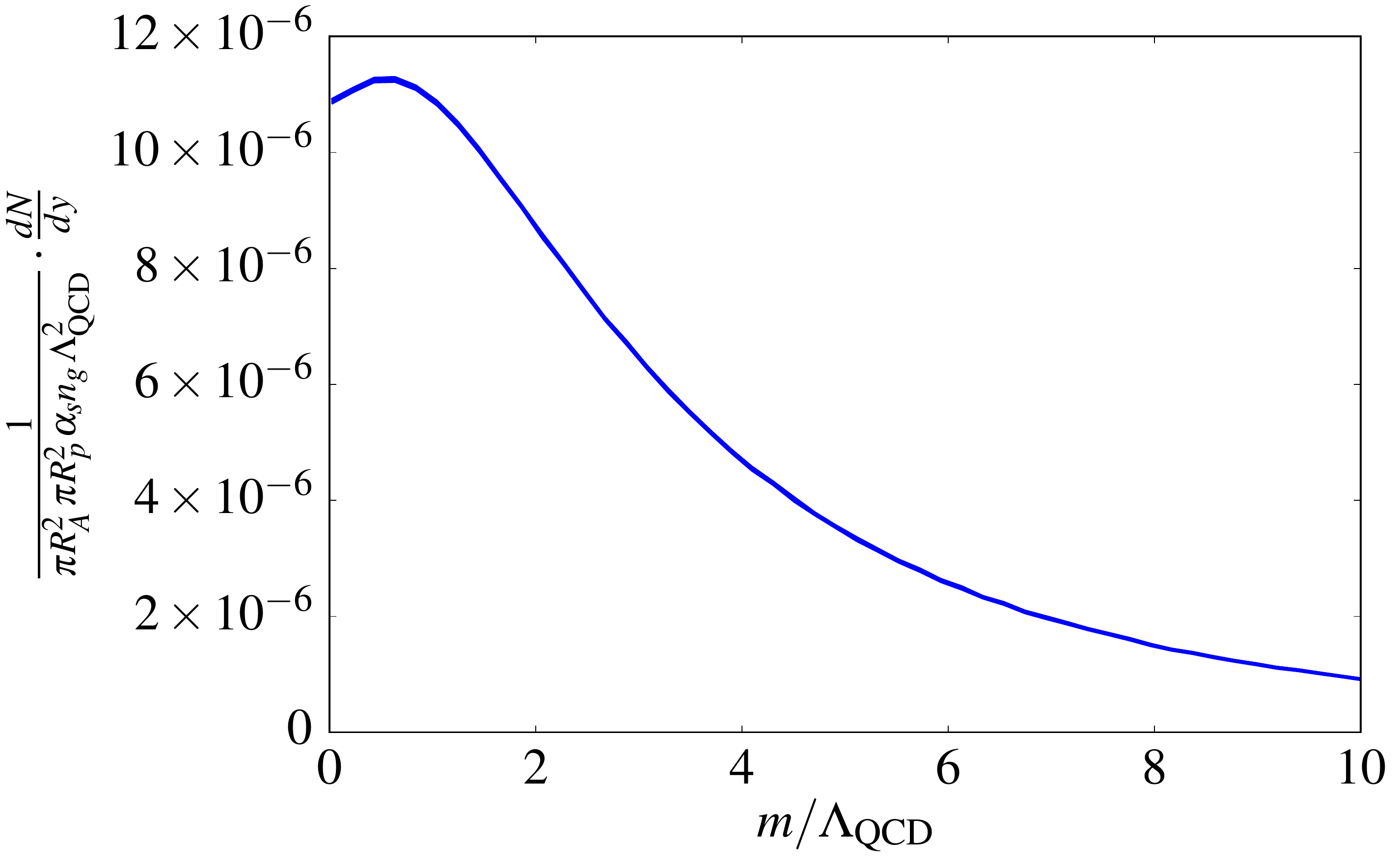}\
 \end{center}
 \caption{Dependence of the total produced photons $dN/dy$ per unit
   rapidity on the quark mass $m$.  The value $Q_s = 10\LQCD$ was
   adopted.  The line thickness represents the $3\sigma$ error
   estimate in the MISER Monte Carlo algorithm.}
\label{fig:nummass}
\end{figure}

According to Eq.~\eqref{eq:num1} the number of produced photons
$dN/dy$ is given as
\be
 \frac{1}{\pi R_A^2 \pi R_p^2 \alpha_s n_g}\cdot \frac{dN}{dy}
 = \frac{\alpha}{2\pi^6}\frac{N_c^2}{N_c^2 - 1}F(0)\;.
\ee 
In Fig.~\ref{fig:nummass} we show $dN/dy$ as a function of the quark
mass for the choice of $Q_s = 10\LQCD$.  We numerically found that the
results for $m\gsim 2\LQCD$ can be well fitted by
$(\log (m/\LQCD))^{1.8}/m^{2.6}$.  From this we can say that the mass
dependence is minor for the strange quark, while the photon production
is suppressed by a factor $\sim 5$ for the charm quark.

\section{Conclusion}
\label{sec:conc}

In this work we have calculated the photon production rate from the
annihilation process with CGC in the $pA$ collision.  We have
considered a regime where the nucleus is saturated and the proton is
more dilute but dominated by gluons over quarks.  We have argued that
in this semi-CGC regime we should consider a process where the virtual
gluon in the proton emits a photon through quark couplings.  We have
obtained an analytic expression for an annihilation part of the
photon rate and identified the part describing saturation physics as
the inelastic quadrupole of Wilson lines.  We have explicitly
demonstrated that the quark loop is both IR and UV finite.  Our
numerical results can be well fitted with a exponential or a
Lorentzian form having a perturbative tail at large momenta.  The
contribution of Bremsstrahlung photons, shown in diagrams (a) and (b)
in Fig.~\ref{fig:LO}, must be explicitly included in the future
phenomenological application.

The multiple scattering effects are included through the MV model but
the small $x$ evolution effects are not yet covered.  However, the
main formula for the photon production rate as given in
Eq.~\eqref{eq:ratf} is quite general and amenable to such systematic
improvements.  In particular, the simple dipole model considered here
could be replaced by the solution of the rcBK equation for future
updates~\cite{Albacete:2007yr}.

To our best knowledge this is the first full numerical integration
over the inelastic quadrupole.  In Ref.~\cite{Kang:2013hta} the
inelastic quadrupole was necessary to predict the singlet quarkonium
production.  However, in the explicit calculations the authors have
used an approximate factorized Ansatz for the quadrupole in terms of
products of dipoles.  Thus, our numerical schemes may have some other
useful applications for related subjects.

Also, the CGC-type quark-loop diagrams, such as the case discussed
here, can in principle be sensitive to the quantum anomaly.  In view
of the possible connection between strong external magnetic fields and
the local parity violation accommodated in the CGC initial state and
the glasma evolution~\cite{Kharzeev:2001ev,Mace:2016svc} our
calculation may be of a broader interest giving a microscopic
foundation of anomaly-induced photons in the early dynamics of the
heavy-ion collision~\cite{Basar:2012bp,Fukushima:2012fg}.  This is one
of intriguing directions for future studies.

For the next step as a continuation from the present work, we plan to
perform a more detailed analysis with physical masses of $u$, $d$,
$s$, and $c$ quarks.  Such a treatment will be indispensable because
quark-loop contributions are sensitive to the explicit breaking of
three ($u$, $d$, $s$) flavor degeneracy.  By including also the
Bremsstrahlung contributions, we can complete the systematic
calculation of photon from CGC in the $pA$ collision and make a full
quantitative prediction/comparison to experimental data.


\section*{Acknowledgments}

We acknowledge discussions with Francois~Gelis. K.~F. thanks Raju~Venugopalan for useful discussions.
S.~B.\ thanks Davor~Horvati\' c and Arata~Yamamoto for their advice on
the numerical routines.
K.~F.\ was supported by MEXT-KAKENHI Grant No.\ 15H03652 and
15K13479.
S.~B.\ was supported by the European Union Seventh Framework Programme
(FP7 2007-2013) under grant agreement No.\ 291823, Marie Curie
FP7-PEOPLE-2011-COFUND NEWFELPRO Grant No.\ 48.
S.~B.\ acknowledges HZZO Grant No.\ 8799 at Zagreb University for
computational resources.

\appendix

\section{Ward identity}
\label{sec:appa}

As an independent check, we explicitly show that the photon amplitude \eqref{eq:mmom} satisfies the Ward identity.
To that end, we must replace $\epsilon_\lambda^\mu(\bk) \to k^\mu$ in the Dirac trace \eqref{eq:dirt1}.
We have
\be
\begin{split}
&{\rm Tr}_D\bigl[\slashchar{k} (\slashchar{q}+m)\slashchar{n}(\slashchar{q}-\slashchar{p}'+m) \bpt\cdot \boldsymbol{\gamma}_\perp(\slashchar{q}-\slashchar{p}-\slashchar{p}' + m)\slashchar{n}(\slashchar{q}-\slashchar{k}+m) \bigr]\\
&= 8p^{+2}\bigl[x p_\perp^2 -2a(x)\bpt\cdot(\bqt-\bpt') \bigr]\bigl[(x\bkt-\bqt)^2+m^2\bigr]\\
&=8p^{+2}\bigl[\frac{1}{2} p_\perp^2 +2a(x)\bpt\cdot\blt \bigr]\bigl[(x\bkt-\bqt)^2+m^2\bigr]~,
\end{split}
\ee
where in the second line we have introduced a substitution $\bpt' \to \blt = \bpt' - \bqt + \bpt/2$ and where $a(x)=x-\frac{1}{2}$.
Defining $\calM_\lambda(\bk) = \epsilon^\mu_\lambda(\bk) \mathcal{T}^\mu(\bk)$ and using the results of Sec.~\ref{ssec:ampl} we have
\be
\begin{split}
k_\mu \mathcal{T}^\mu(\bk) &= \frac{eg}{\pi}\int_0^1 dx\int
 \frac{d^2 \bpt}{(2\pi)^2} \frac{d^2 \blt}{(2\pi)^2} \frac{d^2 \bqt}{(2\pi)^2}
 \,\frac{1}{p_\perp^2}\\
 &\times \frac{1}{(x \bkt-\bqt)^2+m^2}\,\frac{1}{[a(x)\bpt + \blt]^2+m_p^2(x)}\\
  &\times\bigl[\frac{1}{2} p_\perp^2 + 2a(x)\bpt\cdot\blt\bigr]\bigl[(x\bkt-\bqt)^2+m^2\bigr]\\
 &\times \text{Tr}_c \biggl[ U\left(\blt +\bqt - \frac{\bpt}{2}\right)\,
 \rho_p(\bpt)\,U^\dag\left(\blt +\bqt + \frac{\bpt}{2}-\bkt\right)\biggr]\\
&=\int_0^1 dx\int
 \frac{d^2 \bpt}{(2\pi)^2}\frac{d^2 \blt}{(2\pi)^2}
 \,\frac{1}{p_\perp^2} \, \frac{\frac{1}{2} p_\perp^2 + 2a(x)\bpt\cdot\blt}{[a(x)\bpt+\blt]^2+m_p^2(x)}\\
 &\times \int d^2 \byt e^{i(\bpt - \bkt)\cdot\byt}\text{Tr}_c \bigl[ U(\byt)\,
 \rho_p(\bpt)\,U^\dag(\byt)\bigr]\\
& =\frac{\text{Tr}_c[\rho_p(\bkt)]}{k_\perp^2}\int_0^1 dx\int\frac{d^2 \blt}{(2\pi)^2}
 \, \frac{\frac{1}{2} k_\perp^2 + 2a(x)\bkt\cdot\blt}{[a(x)\bkt+\blt]^2+m_k^2(x)}~,
\end{split}
\ee
which vanishes because of the color trace.
In the second line we have recognized that, due to the cancellation between the numerator and the denominator, the dependence of $\bqt$ comes only through the Wilson lines. Fourier transforming the Wilson lines to coordinate space we have performed the $\bqt$ integration. In the third line we have performed the $\byt$ integration.

\section{Angular integrations}
\label{sec:appb}

Here we perform the integration over the polar angle $\phi_p$ associated with the gluon momentum $\bpt$ in Eq.~\eqref{eq:ratf}.
We use the following formulas
\be
\begin{split}
&\int_0^{2\pi} d\phi_p e^{i\bpt \cdot \br} = 2\pi J_0(p_\perp r_\perp)~,\\
&\int_0^{2\pi} d\phi_p \, \bat \cdot \hbp \, e^{i\bpt\cdot \br} = 2\pi i (\bat \cdot \hat{\boldsymbol{r}}_\perp) J_1(p_\perp r_\perp)~,\\
&\int_0^{2\pi} d\phi_p (\bat \cdot \hbp)
(\bbt \cdot \hbp) e^{i\bpt \cdot \br} = 2\pi (\bat \cdot \hbr)
(\bbt \cdot \hat{\boldsymbol{r}}_\perp) J_0(p_\perp r_\perp)\\
&+2\pi \bigl[\bat \cdot \bbt
-2(\bat \cdot \hbr)
(\bbt \cdot \hbr)\bigr]\frac{J_1(p_\perp r_\perp)}{p_\perp r_\perp}~.
\end{split}
\ee
where $J_{0,1}(x)$ are Bessel functions of the zeroth and the first order, respectively.

We define the auxiliary functions
\be
\begin{split}
\varphi_{\mu\nu\rho}^\alpha(x,x',\bu,\bu',\br) &= \int_0^\infty \frac{\varphi(p_\perp) \, p_\perp^{\alpha+2} d p_\perp}{(p_\perp^2+\Lambda_{\rm QCD}^2)^2}\,
m_p^\mu(x) m_p^\nu(x')\\
&\times K_\mu(m_p(x)u_\perp) \, K_\nu(m_p(x')u'_\perp) \, J_\rho(p_\perp r_\perp)~,
\end{split}
\ee 
where $\mu$, $\nu$, $\rho$, $\alpha$ $\in \mathbb{N}_0$.
The angular integration of the first term in Eq.~\eqref{eq:ratf} gives
\be
\begin{split}
&\frac{1}{2\pi}\int d^2 \bpt \frac{p_\perp^4\varphi_p(p_\perp)}{(p_\perp^2 + \LQCD^2)^2}\,
  e^{i\bpt\cdot\br}\Psi_1\Psi'^*_1 = 16 m^2 b(x)b(x')K_1(mu_\perp)K_1(mu'_\perp)\\
&\times\biggl\{a(x)a(x')\varphi_{000}^5+
(\hbu\cdot \hbr)(\hbu'\cdot \hbr)\varphi_{110}^3\\
&+\bigl[\hbu\cdot \hbu'-2(\hbu\cdot \hbr)(\hbu'\cdot \hbr)\bigr]\frac{1}{r_\perp}\varphi_{111}^2-2 a(x')(\hbu'\cdot \hbr)\varphi_{011}^4\biggr\}~,
\end{split}
\ee
where for simplicity we suppressed the variables on $\varphi_{\mu\nu\rho}^\alpha$.
Likewise, the second and the third term in Eq.~\eqref{eq:ratf} are calculated to give
\be
\begin{split}
&\frac{1}{2\pi}\int d^2 \bpt \frac{p_\perp^4\varphi_p(p_\perp)}{(p_\perp^2 + \LQCD^2)^2}\,
  e^{i\bpt\cdot\br}\Psi_2\Psi'^*_2 =m^2 \bigl[K_1(mu_\perp)K_1(mu'_\perp)\varphi_{110}^3\\
&\qquad + m^2 K_0(mu_\perp)K_0(mu'_\perp)\varphi_{000}^3+ 2m K_1(mu_\perp)K_0(mu'_\perp)\varphi_{100}^3\bigr]~,
\end{split}
\ee
\be
\begin{split}
&\frac{1}{2\pi}\int d^2 \bpt \frac{p_\perp^4\varphi_p(p_\perp)}{(p_\perp^2 + \LQCD^2)^2}\,
  e^{i\bpt\cdot\br} \, \hbu \cdot \hbp \Psi_1\Psi'^*_2=4 m^2 b(x)K_1(mu_\perp)\\
&\qquad\times\biggl\{ a(x) \, \hbu\cdot\hbr \bigl[K_1(mu'_\perp)\varphi_{011}^4 + mK_0(mu'_\perp)\varphi_{001}^4\bigr]\\ 
&\qquad-K_1(mu'_\perp)\bigl[(\hbu\cdot \hbr)^2 \varphi_{110}^3+\bigl(1-2(\hbu \cdot\hbr)^2\bigr)\frac{1}{r_\perp}\varphi_{111}^2\bigr]\\
&\qquad-m K_0(mu'_\perp)\bigl[(\hbu \cdot\hbr)^2 \varphi_{100}^3+\bigl(1-2(\hbu \cdot\hbr)^2\bigr)\frac{1}{r_\perp}\varphi_{101}^2\bigr]\biggr\}~.
\end{split}
\ee

\section{Inelastic quadrupole in the MV model}
\label{sec:appc}

The color average of the inelastic quadrupole $S(\by,\bz,\by',\bz')$ (see Eq.~\eqref{eq:trc2}) within the MV model is calculated by the use of the techniques described in  \cite{Blaizot:2004wv,Gelis:2001da,Fukushima:2007dy}.
We closely follow the general procedure described in Ref.~\cite{Fukushima:2007dy} to which we refer the reader for more details. In the notation of \cite{Fukushima:2007dy} the expression for the  
Wilson line product is written as
\be
\begin{split}
&S(\by,\bz,\by',\bz') = \frac{1}{N_c}\big\langle{\rm Tr}\left[U(\by)T_F^a U^\dag(\bz)\right]
{\rm Tr}\left[U(\by')T^a U^\dag(\bz')\right]\big\rangle\\ 
&=\frac{1}{N_c}\delta_{\beta_1\beta_2}\delta_{\beta_3\beta_4}T_{F\alpha_1\alpha_2}^a T_{F\alpha_3\alpha_4}^a
\left\langle U_{\beta_1\alpha_1}(\by)U^*_{\beta_2\alpha_2}(\bz)U_{\beta_3\alpha_3}(\by')U^*_{\beta_4\alpha_4}(\bz')\right\rangle\\
&=\frac{1}{N_c}\delta_{\beta_1\beta_2}\delta_{\beta_3\beta_4}T_{F\alpha_1\alpha_2}^a T_{F\alpha_3\alpha_4}^a \langle \beta_1\beta_2\beta_3\beta_4|e^{-(H_0+V)}|\alpha_1\alpha_2\alpha_3\alpha_4\rangle~,
\label{eq:trc}
\end{split}
\ee
where \cite{Fukushima:2007dy}
\be
H_0 = \frac{2N_c}{N_c^2-1}(T_{F1}^a - T_{F2}^{a*}+T_{F3}^a - T_{F4}^{a*})^2 L(0)~,
\ee
\be
\begin{split}
V = -\frac{2N_c}{N_c^2-1} \bigl[&-T_{F2}^{a*} T_{F1}^a B_2(|\by-\bz|)+T_{F3}^a T_{F1}^a B_2(|\by-\by'|)\\
&-T_{F3}^a T_{F2}^{a*} B_2(|\bz-\by'|)-T_{F4}^{a*} T_{F1}^a B_2(|\by-\bz'|)\\
&+T_{F4}^{a*} T_{F2}^{a*} B_2(|\bz-\bz'|)-T_{F4}^{a*} T_{F3}^a B_2(|\by'-\bz'|)\bigr]~.
\end{split}
\ee
Here
\be
L(|\bx - \by|) = Q_s^2\int d^2 \bz G_0(\bx-\bz)G_0(\by-\bz)~,
\ee
with $G_0(\bx)$ a Green function of a $2D$ Laplacian 
\be
\partial_{\bx}^2 G_0(\bx) = \delta^{(2)}(\bx)~,
\ee
and
\be
B_2(|\bx - \by|) = 2L(0)-2L(\bx-\by)~.
\ee

Define now $|\alpha\rangle \equiv|\alpha_1\alpha_2\alpha_3\alpha_4\rangle$.
The Wilson line product has two singlets, defined as
\be
\langle \alpha|s_1\rangle = \frac{1}{N_c}\delta_{\alpha_1\alpha_2}\delta_{\alpha_3\alpha_4}~,\qquad
\langle \alpha|s_2\rangle = \frac{2}{\sqrt{N_c^2-1}}T_{F\alpha_1\alpha_2}^a T_{F\alpha_3\alpha_4}^a~,
\label{eq:s12}
\ee
Note that precisely these singlets appear in the contraction of the Wilson line product (\ref{eq:trc}).
This leads to
\be
\begin{split}
S(\by,\bz,\by',\bz') &= \frac{\sqrt{N_c^2-1}}{2}\sum_{\alpha,\beta}\langle s_1 |\beta\rangle \langle \beta | e^{-(H_0+V)} |\alpha\rangle \langle \alpha | s_2 \rangle\\ 
& = \frac{\sqrt{N_c^2-1}}{2}\langle s_1 | e^{-V} |s_2\rangle~,
\end{split}
\label{eq:trs}
\ee
where in the first line we used the definition of the singlets \eqref{eq:s12}.
In the second line we have used $H_0|s_{1,2}\rangle = 0$. 
From \cite{Fukushima:2007dy} (see also \cite{Blaizot:2004wv}), we have
\be
\langle s_1| e^{-V}|s_2\rangle  =-\frac{2 V_{12}}{\phi}e^{-\frac{1}{2}{\rm Tr}V} \sinh\frac{\phi}{2}~,
\label{eq:int12}
\ee
where the $2\times 2$ matrix $V_{ij} = \langle s_i |V|s_j\rangle$ is given as \cite{Fukushima:2007dy}
\be
\begin{pmatrix}
V_{11} & V_{12}\\
V_{21} & V_{22}
\end{pmatrix}
= - \frac{2 N_c}{N_c^2-1}
\begin{pmatrix}
-\frac{N_c^2-1}{N_c}\gamma & \frac{\sqrt{N_c^2-1}}{N_c}(\beta-\alpha)\\
\frac{\sqrt{N_c^2-1}}{N_c}(\beta-\alpha) & \frac{1}{N_c}(\gamma-2\beta + (2-N_c^2)\alpha)
\end{pmatrix}
~.
\label{eq:vmat}
\ee The functions $\alpha$, $\beta$ and $\gamma$ were defined in Eq.~\eqref{eq:abg} and $\phi = \sqrt{({\rm tr} V)^2  - 4{\rm det}V}$.
A simple algebraic manipulation of \eqref{eq:int12} leads  to
\begin{equation}
 \begin{split}
 S(\by,\bz,\by',\bz') &= \frac{1}{N_c}
 \frac{(N_c^2-1) (\beta-\alpha)}{\sqrt{N_c^2 (\alpha-\gamma)^2
 -4(\alpha-\beta)(\beta-\gamma)}}\\
 &\times \exp\biggl[-\frac{2\beta+(\alpha+\gamma)(N_c^2-2)}{N_c^2-1}\biggr]\\
 &\times\sinh\biggl[\frac{N_c}{N_c^2-1}
 \sqrt{N_c^2 (\alpha-\gamma)^2-4(\alpha-\beta)(\beta-\gamma)}\biggr]\;.
\end{split}
\end{equation}


\end{document}